\documentclass[useAMS,usenatbib,fleqn]{mn2e}
\usepackage{subfigure}
\usepackage{graphicx, amssymb}
\usepackage[fleqn]{amsmath}
\usepackage{epsfig}
\usepackage{color}
\usepackage{times}
\usepackage{epstopdf}

\title[Star Cluster Formation in the First Galaxies]{Star formation in the first galaxies - III. Formation, evolution, and characteristics of the first stellar cluster}

\author[C. Safranek-Shrader et al.]
{Chalence~Safranek-Shrader$^{1,2}$, Michael~H.~Montgomery$^{2,3}$,  Milo\v s~Milosavljevi\'c$^{2}$,\newauthor and Volker~Bromm$^{2}$\\
$^{1}$Department of Astronomy and Astrophysics, University of California, Santa Cruz, CA 95064, USA \\
$^2$Department of Astronomy, University of Texas at Austin, TX 78712, USA \\
$^3$McDonald Observatory, University of Texas at Austin, TX 78712, USA 
}
 
\newcommand{\nh}{n_{\mathrm{H}}}
\newcommand{\kelvin}{\mathrm{K}}
\newcommand{\cc}{\mathrm{cm}^{-3}}
\newcommand{\msun}{M_{\odot}}

\newcommand{\zsun}{Z_{\odot}}
\newcommand{\zcrit}{Z_{\mathrm{crit}}}
\newcommand{\htwo}{\mathrm{H}_2}

\newcommand{\tcmb}{T_{\mathrm{CMB}}}

\newcommand{\tff}{t_{\mathrm{ff}}}

\newcommand{\kb}{k_{\mathrm{B}}}
\newcommand{\mh}{m_{\mathrm{H}}}

\newcommand{\pc}{\mathrm{pc}}
\newcommand{\au}{\mathrm{AU}}

\newcommand{\tvir}{T_{\mathrm{vir}}}
\newcommand{\mvir}{M_{\mathrm{vir}}}
\newcommand{\jlw}{J_{\mathrm{LW},21}}
\newcommand{\cs}{c_{\mathrm{s}}}

\newcommand{\lj}{L_{\mathrm{J}}}

\newcommand{\microm}{\mu{\rm m}}

\newcommand{\racc}{r_{\mathrm{acc}}}
\newcommand{\msunperyr}{M_{\odot}\,\mathrm{yr}^{-1}}

\newcommand{\tdust}{T_{\mathrm{d}}}
\newcommand{\betaesc}{\beta_{\mathrm{esc}}}

\newcommand{\taucont}{\tau_{\mathrm{cont}}}

\begin{document}

\label{firstpage}

\maketitle
\topmargin-1cm

\begin{abstract}
We simulate the formation of a low metallicity ($10^{-2}\,\zsun$) stellar cluster in a dwarf galaxy at redshift $z\sim 14$. Beginning with cosmological initial conditions, the simulation utilizes adaptive mesh refinement and sink particles to follow the collapse and evolution of gas past the opacity limit for fragmentation, thus resolving the formation of individual protostellar cores.  A time- and location-dependent protostellar radiation field, which heats the gas by absorption on dust, is computed  by integration of protostellar evolutionary tracks with the \textsc{mesa} code. The simulation also includes a robust non-equilibrium chemical network that self-consistently treats gas thermodynamics and dust-gas coupling. The system is evolved for $18$ kyr after the first protostellar source has formed. In this time span, $30$ sink particles representing protostellar cores form with a total mass of $81\,\msun$.  Their masses range from $\sim0.1\,\msun$ to $14.4\,\msun$ with a median mass $\sim0.5-1\,\msun$.  Massive protostars grow by competitive accretion while lower-mass protostars are stunted in growth by close encounters and many-body ejections. In the regime explored here, the characteristic mass scale is determined by the temperature floor set by the cosmic microwave background and by the onset of efficient dust-gas coupling. It seems unlikely that host galaxies of the first bursts of metal-enriched star formation will be detectable with the James Webb Space Telescope or other next-generation infrared observatories. Instead, the most promising access route to the dawn of cosmic star formation may lie in the scrutiny of metal-poor, ancient stellar populations in the Galactic neighborhood. The observable targets that correspond to the system simulated here are ultra-faint dwarf satellite galaxies such as Bo\"otes II, Segue I and II, and Willman I.
\end{abstract}


\section{Introduction}
\label{sec:intro}

The cosmic `dark ages' was a transformative epoch in the history of the Universe, witnessing the birth of the first stars, galaxies, and supermassive black holes, the beginning of reionization, and the onset of chemical enrichment \citep{Barkana01}. Despite its importance, it is also one of the least understood epochs.  Metal-poor stars and dwarf galaxies, likely relics from the high-redshift Universe, present an opportunity to use local observations to investigate the earliest stellar generations and their role in driving early cosmic evolution \citep[e.g.,][]{Beers05,Frebel12,Karlsson13,Frebel13}. To capitalize on the power of stellar and the Galactic dwarf satellite galaxy `archaeology' as a probe of primeval stellar populations, it is necessary to develop a basic theoretical understanding of star formation as it occurred in the first galaxies at high cosmic redshifts and low metallicites.


Modeling star formation in the first galaxies is challenging in part because we still lack a robust, predictive theory of present-day star formation \citep[see reviews by][]{McKee07,Zinnecker07,Krumholz14_protostars_planets,Tan14}, let alone for that occurring in the early Universe \citep{Bromm13}. Nevertheless, one plausible conjecture is that gas metallicity, likely highly sub-solar in the first galaxies \citep[e.g.,][]{Wise08,Greif10}, moderated the thermodynamics of star forming gas \citep[e.g.,][]{Larson85}.

The thermodynamic evolution of collapsing gas concentrations, determined through the competition of heating and cooling processes, is known to be crucial for modulating fragmentation \citep[e.g.,][]{Li03} and for potentially fixing the characteristic (i.e., median) mass of the stellar initial mass function \citep[IMF; e.g.,][]{Masunaga98,Omukai00,Larson05,Jappsen05}. In present-day molecular clouds, the primary heating sources are dust grain photoelectric absorption and Coulomb collisions with cosmic rays. Cooling occurs through line emission (C$^+$ or CO) and dust grain thermal emission, at low and high densities, respectively. While the precise balance of these processes depends largely on the metallicity, column density, and strength of external radiation fields, star forming gas evolves approximately isothermally until the column density has increased enough for the dust to become opaque to its own cooling radiation.  After that, the evolution is quasi-adiabatic. 

In Population III (Pop III) stars which form from metal-free gas, the thermodynamic evolution is markedly different, dictated instead by the physics of molecular hydrogen, $\htwo$, the most effective coolant in low-temperature ($T<10^4\,\kelvin$), metal-fee gas \cite[e.g.,][]{Silk83,Abel00,Bromm02,Yoshida06}. This is believed to make the characteristic Pop III stellar mass larger \citep[e.g.,][]{Stacy10,Clark11,Greif11,Hirano14} than the typical Galactic stellar mass of $\sim0.1\,\msun$ \citep[e.g.,][]{Chabrier03}.


The salient heating and cooling processes at temperatures below $\sim10^4\,\kelvin$ are sensitive to metallicity. Using radiation hydrodynamical (RHD) simulations, \citet{Myers11} and \citet{Bate14} examined the effect of metallicity variation on the star formation process in the context of present-day star formation. They found that down to $10^{-2}\,\zsun$, metallicity had little effect on the IMF and other properties of simulated stellar systems, in agreement with astronomical findings \citep[e.g.,][]{Bastian10}. In these studies, though, dust and gas were assumed to be perfectly collisionally coupled, so that the sole effect of lowering metallicity was a proportional reduction in dust opacity. In reality, the assumption of perfect dust-gas coupling breaks at low densities and metallicities, an effect that can significantly alter the gas cooling rate and thermodynamic behavior.

The impact of metals on the thermodynamic evolution of collapsing gas concentrations has also been studied in the theoretical context of the Pop III to Pop II star formation mode transition. Idealized, one-zone thermodynamic models generally suggest that once the gas metallicity has exceeded some critical value $\zcrit$, the thermodynamical behavior, and thus characteristic fragmentation mass, change drastically, either due to metal line cooling \citep[$\zcrit\sim10^{-3.5}\,\zsun$; e.g.,][]{Bromm03,Santoro06,SafranekShrader10,Omukai10}, or due to the onset of gas-dust collisional coupling \citep[$\zcrit\sim10^{-6}\,\zsun$; e.g.,][]{Omukai05,Schneider06,Schneider10}. Multidimensional hydrodynamic simulations have largely confirmed this
\citep{Bromm01,Smith07,Clark08,Smith09,Dopcke11,Dopcke13, SafranekShrader14a,SafranekShrader14,Bovino14,Meece14}.

A thorough understanding of star formation in the first galaxies requires placing the process in its proper cosmological context. The first galaxies are anticipated to form in dark matter halos with virial temperatures  $\tvir\sim10^4\,\kelvin$ corresponding to virial masses $\mvir\sim10^8\,[(1+z)/10]^{-3/2}\,\msun$ \citep{Oh02,Bromm11}. In these `atomic cooling halos', energetic supernovae, infall of baryonic matter from the cosmic web, and perhaps even thermal instability, all contribute to the excitation of turbulence in the star-forming gas \citep[ISM;][]{Wise07,Greif08,Wise08a,Prieto12,SafranekShrader12}. Previous stellar generations could have  polluted these halos to average metallicities $Z>\zcrit$ \citep[e.g.,][]{Tornatore07,Wise08,Greif10,Maio10,Ritter12}, which would have enabled the formation of Pop II stars \citep[but metal pollution to $Z>\zcrit$ cannot be taken for granted, see][]{Ritter14}.

If metal enriched star formation promptly followed Pop III star formation \citep[e.g.,][]{Ricotti02,Ricotti08,Ritter12,Whalen13}, $\mvir \sim10^6\,\msun$ `minihalos' would have been the host sites of the very first Pop II stellar clusters. A minihalo, however, does not meet other intuitive criteria for a \emph{bona fide} galaxy, such as the requirement that it be able to retain photoionized gas. This taxonomic subtlety aside, the physical state of primordial star forming systems just prior to the onset of metal-enriched star formation has been explored in a number of studies, all reaching largely compatible conclusions. This physical state defines the initial conditions for metal-enriched star formation in the first galaxies. 

Simulations of the transformation of a gas cloud into stars are traditionally initialized with an isolated spherical or quasi-spherical overdensity. A spectrum of random velocity perturbations is imposed at initialization, or hydrodynamic turbulence is externally driven until saturation. The parameters of the initial cloud and of the velocity (or perturbing force) fluctuations are often chosen ad hoc, reflecting a focus on the rudimentary mechanisms of star formation, rather than on making quantitative predictions in  varied, realistic, complex astrophysical settings. Properties of the resulting stellar systems, such as their IMFs \citep{Klessen01,Martel06,Clark08,Urban10,Girichidis11}, or the mode of the stellar mass fixation \citep[e.g., `core accretion' vs.\ `competitive accretion', see][]{Krumholz05,Bonnell06}, exhibit sensitivity to the model parameters. Encouragingly, state-of-the-art radiation-hydrodynamical (RHD) simulations have successfully delivered stellar clusters with convergent IMFs and other statistical measures matching observed low-luminosity star forming regions such as the Orion nebula cluster \citep[e.g.,][]{Krumholz12,Bate12}, though fine tuning of certain parameters is still required. Star formation simulations that begin from \emph{ab initio} initial conditions---either seeded from global galactic disk simulations, or, ultimately, by inflationary random fields---are needed to settle these uncertainties.

In the first simulation to deliver low-mass protostars directly from cosmological random fields \citep{SafranekShrader14a}, we simulated the assembly of a metal-poor ($10^{-2}\,\zsun$) stellar system in a redshift $z\sim14$ atomic cooling halo. The initial conditions were excised from the cosmological box of \citet{SafranekShrader14} evolved to peak densities in gravitationally collapsing gas clumps of $\approx 10^7\,\cc$ . The excised box was centered on a single dense clump that formed via metal-line-cooling-induced thermal instability.  We adaptively refined the computational grid and introduced Lagrangian sink particles to track gravitational collapse of gas to densities $\sim 10^{13}\,\cc$ and resolved the formation of \emph{individual} protostellar cores. Over the maximum simulation time of $\approx7\,\mathrm{kyr}$ permitted by computing resources, $40$ protostellar cores formed with final masses ranging from $10^{-3}\,\msun$ to $2.5\,\msun$. The stellar IMF above $0.1\,\msun$ was tentatively shallower (top-heavier) than the Salpeter power law. Heating by protostellar accretion radiation did not have an impact on the low-mass IMF, but the time extent of the simulation following the onset of star formation, corresponding to only a tenth of the local free fall time, was insufficient for massive, luminous protostars to form and appreciably heat the dust. In principle, we could have integrated the system as much as $\sim10$ times longer while still avoiding boundary artifacts from the box excision. Such an extension of the simulation would have allowed the protostellar mass function to grow to higher masses and much higher luminosities.  

In the present paper, we extend the simulation of \citet{SafranekShrader14a} while crucially calibrating individual protostellar luminosities to evolutionary tracks computed with the new stellar evolutionary package {\it Modules for Experiments in Stellar Astrophysics} \citep[\textsc{mesa};][]{Paxton11,Paxton13}. In addition to these improvements, we provide the technical detail left out in the \emph{Letter}-length report of our early results in \citet{SafranekShrader14a}.

We organize this paper as follows. In Section \ref{sec:method} we describe the methodology and simulation setup. This includes a brief description for the parent cosmological simulation from which our initial conditions were excised as well as a detailed description of our chemo-thermodynamic and protostellar evolutionary model. In Section \ref{sec:results} we describe the results of the simulation with a special emphasis on the formation and growth of sink particles and the thermodynamics of the gas and dust in the course of gravitational collapse. In Section \ref{sec:discussion} we discuss the implications of our results and summarize our conclusions.

\section{Methodology}
\label{sec:method}


\subsection{Simulation setup}
\label{sec:setup}

We performed our simulation with the adaptive-mesh-refinement (AMR) hydrodynamics code
\textsc{flash} \citep{Fryxell00}, version 4. The initial conditions were extracted from the cosmological simulation of \citet{SafranekShrader14} and consisted of a cubical region of size
$0.52\,\pc$ containing a total gas mass of $390\,\msun$. In this parent simulation, a cosmological volume of $1$ comoving Mpc$^3$ was evolved until an atomic cooling halo with a virial temperature of $\tvir \approx 10^4\,\kelvin$ formed in the box. Emulating a radiative background due to star formation outside the cosmological box, we introduced a global Lyman-Werner (LW) radiation field with intensity $\jlw=100$ which had the effect of photodissociating $\htwo$ in low-column-density gas and preventing Pop III star formation in the progenitor minihalos.\footnote{Here, $J_{{\rm LW},21}$ is the mean intensity at the center of the LW bands, $12.4 \,{\rm eV}$, expressed in the units of $10^{-21}\,\mathrm{erg}\,\mathrm{s}^{-1}\,\mathrm{cm}^{-2}\,\mathrm{Hz}^{-1}\,\mathrm{sr}^{-1}$. This is closely related, thought not exactly equal, to $J_{21}$, the radiation intensity at the Lyman limit.} The atomic cooling halo virialized with mass  $M_{\rm vir}\approx 2\times10^7\,\msun$ at $z=13.8$.  At that point, we set the gas metallicity inside its virial radius from zero to $10^{-2}\,\zsun$, modeling metal enrichment by preceding Pop~III stars in a highly idealized fashion.  The excision of a dense clump was performed when the peak density reached $10^7\,\cc$. The gravitational potential in the excised region was strongly baryon-dominated and so after excision we neglected dark matter. The mass-weighted average density inside the box was $9\times10^5\,\cc$ corresponding to a free-fall time of $\tff = (3\pi / 32G\rho)^{1/2}\approx 50\,\mathrm{kyr}$. 

The excised hydrodynamic data cube was used to refine the simulation to densities and time scales unattainable in the larger cosmological box while still retaining direct causal dependence on the cosmological random fluctuations. We imposed reflective boundary conditions crudely approximating a pressure-confined environment. To prevent contamination of our results by boundary effects, we could run the simulation only for much shorter than the sound crossing time from box edge to center, $\sim 0.4\,\mathrm{Myr}$. We always resolved the Jeans length
 \begin{equation}
\lj = \left(\frac{\pi \cs^2}{G\rho}\right)^{1/2} 
\label{eq:lj}
\end{equation}
by at least 24 grid cells. Through experimentation we found that resolving the Jeans length by fewer cells produced a significantly different turbulent morphology and non-convergent protostellar growth trends (see, also, \citealt{Federrath11} and \citealt{Turk12}).




\subsection{Sink particles and the formation of individual stars}
\label{sec:sink}

Above some critical column density, collapsing gas concentrations generally transition from isothermal to adiabatic evolution when the continuum opacity exceeds unity and radiative cooling loses efficacy. This is the point when the energy input rate from gravitational compression exceeds the maximum allowed radiative loss rate. The thermal Jeans mass at the isothermal-to-adiabatic transition is known as the opacity limit for fragmentation \citep{Rees76,Low76}. It represents the minimum mass of a gravitationally collapsing gas clump.
\citet{Omukai00} showed that if
the optical depth is estimated
as the opacity multiplied with the local Jeans length, as is appropriate in
self-gravitating collapsing gas, the density at which the optical depth equals unity
is independent of opacity (and thus of metallicity) and is given by 
\begin{eqnarray}
n_{\mathrm{crit}} &=& \left(\frac{12\,T^5 \sigma_{\rm SB}^2\mh}{\kb^3}\right)^{1/2} \nonumber\\&\approx& 5\times10^{11}\,\cc\,\left(\frac{T}{40\,\kelvin}\right)^{5/2} \, ,
\label{eq:n_tau}
\end{eqnarray}
where $T$ is the gas temperature, $\mh$ is the hydrogen atom mass, and $\kb$ and $\sigma_{\rm SB}$ are the Boltzmann and Stefan-Boltzmann constants, respectively. Thus, to resolve the formation of individual stars, we must follow the gas collapse past $n_{\mathrm{crit}}$.

Sink particles (hereafter sinks) are a commonly employed in simulations of star formation. Originally introduced by \citet{Bate95}, they have been widely used in both Eulerian (Godunov-type) and Lagrangian (smoothed particle hydrodynamics; SPH) methods. As self-gravitating gas collapses and density increases, the Eulerian grid refines itself to resolve the local Jeans length by a minimum number of resolution elements.  Eventually, the Courant-Friedrichs-Lewy limited time step becomes prohibitively short. To continue the simulation past the initial runaway collapse, it becomes necessary to introduce a subgrid model for collapsed sites.  This is done by inserting Lagrangian, collisionless, gravitating particles into sites where the density has exceeded a threshold. The particles accrete excess density gas from the grid in a momentum-conserving fashion, keeping the Jeans length in check and obviating further grid refinement. 

Sinks, however, are not just a computational shortcut.  If used correctly, they can represent distinct, localized, gravitationally collapsed gas concentrations.  They make it straightforward to record the mass accretion rate and other subgrid features of the histories of these concentrations. To regard sinks as \emph{individual} protostellar cores (and minimize the chance of unresolved sub-fragmentation occurring within a sink), we set the density at which sinks form to $n_{\mathrm{sink}}=4\times10^{12}\,\cc$, nearly an order of magnitude higher than the estimate given by Equation (\ref{eq:n_tau}). Physically, this means that gas that reaches $n_{\mathrm{sink}}$ becomes incorporated in a well-defined, pressure-supported, quasi-adiabatic core in which gravitational fragmentation is suppressed. 

In addition to the density threshold, for sink particle creation we also require that the 
flow in the cell with $n>n_{\mathrm{sink}}$ be converging $\nabla\cdot\textbf{v}<0$, the
gravitational potential be a local minimum, and a small control
volume around the cell be gravitationally bound, $E_{\mathrm{therm}} + E_{\mathrm{kin}} + E_{\mathrm{grav}}<0$. Here, $E_{\mathrm{kin}}$ is the total kinetic energy evaluated with respect to the center-of-mass velocity.  These additional checks are essential to prevent the formation of spurious sink particles that do not represent physical gravitationally collapsing sites \citep{Federrath10}.  In each hydrodynamic time step, grid cells within the sink's accretion radius of $\racc = 20\,\au =
 2.5\,\Delta x_{\mathrm{min}}$ with densities of hydrogen nuclei
 $\nh>n_{\mathrm{sink}}$ transfered a fraction
 $(\nh-n_{\mathrm{sink}})/\nh$ of their mass to the sink if the gas
 was gravitationally bound to the sink and had a  velocity
 with a negative radial component relative to the sink. 
 Here, $\Delta x_{\rm min}$ is the cell size at
 the highest level of grid refinement. Sink particle motion is sub-cycled with a leapfrog integration scheme. Sinks are not allowed to merge with each other. For further details and limitations of the sink particle approach, see \citet{Federrath10}.


%

 \subsection{Thermodynamical model}
 \label{sec:thermo}
 
Our thermodynamical model and non-equilibrium chemical network were as in \citet{SafranekShrader10,SafranekShrader12,SafranekShrader14}, 
but now with the dust processes described in \citet{Omukai05} and additional processes that become relevant at high densities, $\nh\gtrsim10^8\,\cc$. At sub-solar metallicities and low densities, dust and gas cannot be assumed to be thermally coupled. In general, their temperatures differ.  We self-consistently calculated the dust temperature $\tdust$ by assuming balance between cosmic microwave background (CMB) and protostellar radiation absorption, thermal emission, and inelastic gas-dust collisions.  We numerically solved the balance relation
 \begin{align}
 4\sigma_{\rm SB}(\tdust^4 - \tcmb^4)\kappa_{\rm d}&(\tdust)\rho \betaesc = \frac{2\kb (T-\tdust)n_{\mathrm{d}}}{t_{\mathrm{coll}}} \nonumber \\
 &+ \sum_i\left(\frac{L_{\mathrm{tot},i} }{4\pi r^2_i}\right)\,\kappa_{\rm d}(\tdust)\rho\betaesc \,
 \label{eq:tdust_eqn}
\end{align}
for the dust temperature $\tdust$.
Here, $\rho$ is the gas density, $T_{\rm CMB}$
is the CMB temperature, $r_i$ is the distance to the $i$th sink, and $L_{\mathrm{tot},i}$ is the total protostellar luminosity of the $i$th sink (see Section \ref{sec:protostellar}). We assumed that the number density of dust grains scaled linearly with metallicity $n_{\mathrm{d}}=\rho D (Z/\zsun) / m_{\rm d}$, where $D$ is the Galactic dust-to-gas ratio, which we took to be $D=0.01$, and $m_{\rm d}=1.3\times10^{-14}\,{\rm g}$ is the effective dust grain mass \citep{Cazaux04}. The collision time
between gas and dust particles is $t_{\mathrm{coll}}^{-1} =
\nh\sigma_{\mathrm{d}}\bar{v}_{\mathrm{H}} f$, where
$\sigma_{\mathrm{d}}=0.1\,\mu {\rm m}$ is the average dust grain cross section \citep{Cazaux04},
$\bar{v}_{\mathrm{H}}$ is the average speed of hydrogen nuclei, and
$f\approx0.4$ accounts for non-hydrogenic species
\citep{Schneider06}. The Planck mean opacity of dust grains,
$\kappa_{\rm d}(\tdust)$, was taken from \citet{Semenov03}. We assumed
that $\kappa_{\rm d}(\tdust)$ scales linearly with metallicity and adopted a density-independent dust sublimation temperature of $1500\,\kelvin$. Thermal emission from dust grains is attenuated by
a factor $\betaesc = \mathrm{min}(1,\taucont^{-2})$, appropriate in the regime of 
optically-thick radiative diffusion \citep[e.g.,][]{Masunaga98}. The
continuum optical depth is given by $\taucont = (\kappa_{\rm d} +
\kappa_{\rm g})\rho\lj$, where $\lj$ is used as a
local estimate of the physical extent of a gravitationally collapsing core and 
$\kappa_{\rm g}(\rho,T)$ is gas Planck mean opacity.
We ignored the metal contribution to $\kappa_{\rm g}$ and took the gas opacity from \citet{Mayer05}.  Ignoring metals in $\kappa_{\rm g}$ is justified because the bulk of metal opacity is accounted for in $\kappa_{\rm d}$. 

To determine the metal fine structure line cooling rate, we utilized the same procedure as in \citet{SafranekShrader14}, but now model radiation trapping by line and dust opacity. Since the level populations and line escape probabilities depend on each other, obtaining a self-consistent cooling rate requires iterative solution \citep[e.g.,][]{Takahashi83,Omukai00}. We employed a local estimate of the Sobolev length,
\begin{equation}
\label{eq:Sobolev}
L_{\mathrm{sob}} = \frac{\cs}{
|\nabla\cdot\textbf{v}|} \,, 
\end{equation}
to approximate the size of the
shielding region.

Our computations accounted for cooling by ro-vibrational lines of $\htwo$.  Molecular hydrogen forms via 
the $\mathrm{H}^-$ intermediary and on the surfaces of dust grains.  Above $n\sim10^8\,\cc$, $\htwo$ begins to form efficiently in three-body reactions.  We track the latent heat of $\htwo$ formation and dissociation. The reduction in the ro-vibrational $\htwo$ cooling rate due to line trapping
  was modeled using an escape probability formulation based on the same local estimate, Equation (\ref{eq:Sobolev}), of the Sobolev length \citep[see, e.g.,][]{Yoshida06}.\footnote{This approach may overestimate the $\htwo$ line emission escape fraction by a large factor in certain regimes \citep{Greif14}.} Our model also included collisionally induced emission (CIE) from 
  $\htwo$, though this process does not become a significant cooling channel
  until densities exceed $\sim10^{14}\,\cc$, a regime not explored here.

  \begin{figure}
\begin{center}
\includegraphics[width=0.48\textwidth]{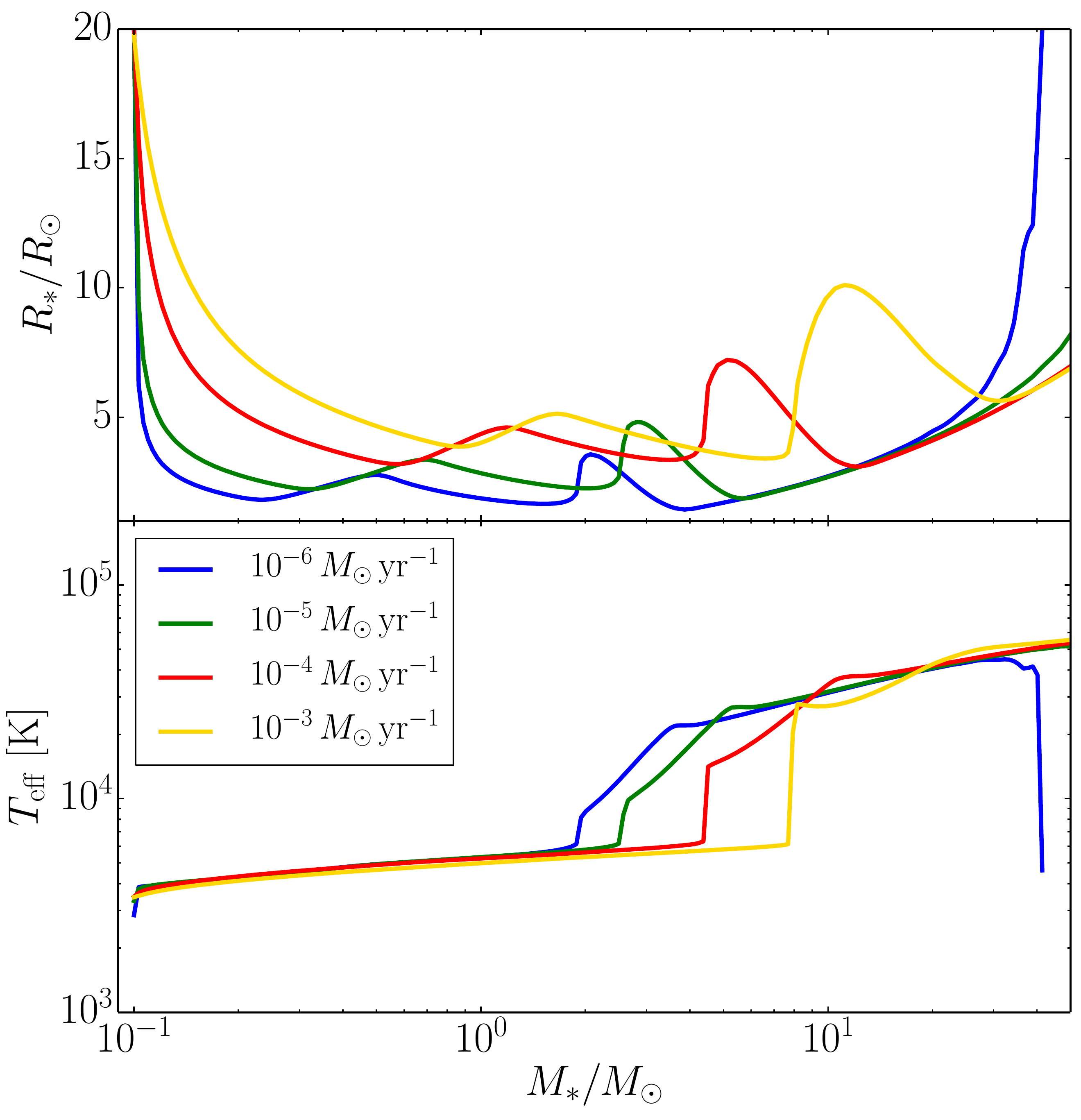}
\end{center}
\caption{Protostellar radius (top panel) and effective temperature (bottom) as a function of stellar mass from our \textsc{mesa} evolutionary tracks of accreting protostars. The colors refer to different values of the constant mass accretion rate (see legend). Deuterium burning is active at $M_*\sim (0.3-0.5)\,\msun$ for $\dot{M}_* = 10^{-6}\,\msun\,\mathrm{yr}^{-1}$ and at $M_*\sim (0.5-2)\,\msun$ for $\dot{M}_* = 10^{-3}\,\msun\,\mathrm{yr}^{-1}$ and is manifested in a shallow bump in the stellar radius. Radius and the effective temperature show a notable, sudden increase with core contraction and the associated convective-to-radiative transition.  Kelvin-Helmholtz contraction toward the zero-age main sequence (ZAMS) follows thereafter. Above $\sim20\,\msun$, the $10^{-6}\,\msunperyr$ track begins ascent of the red giant branch.  \label{fig:mesa}}
\end{figure}

\subsection{Protostellar growth and feedback}
\label{sec:protostellar}

Radiation emitted from the surfaces and disks of accreting protostars can modify conditions in the
ambient star-forming gas \citep[e.g.,][]{Offner09}.
The summand in the last term of Equation (\ref{eq:tdust_eqn}) represents the heating rate of dust by
the radiation of the $i\mathrm{th}$ protostar located at distance $r_i$ with a total luminosity 
\begin{equation}
L_{\mathrm{tot}} = L_{\mathrm{int}} + L_{\mathrm{acc}}
\end{equation}
that is a combination of the
 intrinsic luminosity 
 \begin{equation}
L_{\mathrm{int}} = 4\pi R_{*}^2 \sigma_{\rm SB} T_{\mathrm{eff}}^4 \, ,
\label{eq:l_surf}
\end{equation}
and accretion luminosity
 \begin{equation}
L_{\mathrm{acc}} = f_{\mathrm{acc}} \frac{GM_{*}\dot{M}_{*}}{R_{*}}\, ,
\label{eq:l_acc}
\end{equation}
where $f_{\rm acc}$ is a dimensionless coefficient quantifying radiative efficiency of accretion.
By treating sinks as sources of radiation, we identify $M_*$ with the sink
mass and $\dot{M}_*$ with rate of accretion onto the sink.  The accretion rate was
smoothed over a $2$ yr window corresponding to about $5$ hydrodynamical time steps. We assume a fixed radiative efficiency of $f_{\mathrm{acc}}=0.75$ \citep[e.g.,][]{Offner09}. This choice hides our ignorance of the detailed geometry of the protostellar accretion flow, as we explain below.

The radius of the protostellar photosphere $R_*$ and the effective temperature $T_{\rm eff}$ evolve significantly over the course of protostellar growth. We used \textsc{mesa} to tabulate the protostellar radius and effective temperature as a function of protostellar mass and accretion rate.  Each \textsc{mesa} track was computed with a constant (time-independent) mass accretion rate $\dot{M}_*$.  The entropy of the accreted gas was set to that of the stellar surface, approximating accretion via a protostellar disk. We assumed standard Big Bang nucleosynthesis (BBN) abundances for hydrogen and helium, a metallicity of $10^{-2} Z_\odot$, and the standard solar abundance pattern \citep{Grevesse98}. For $T_{\rm eff} < 43,000\,\kelvin$, \textsc{mesa} photospheric tables were used as the atmospheric boundary condition, while for $T_{\rm eff}>43,000\,\kelvin$, an Eddington grey atmosphere was assumed. For deuterium burning we used the prescription of \citet{Weiss04} for convection with a mixing-length-to-pressure-scale-height ratio of $\alpha_{\rm ML} = 2.0$. The \textsc{mesa} integrations were initialized with $M_* = 0.1 \,\msun$ and $R_* = 20\, R_\odot$ and  stopped when either the stellar mass reached $50\, M_\odot$ or the star left the main sequence. In Figure \ref{fig:mesa} we show the protostellar radius and effective temperature as a function of protostellar mass for four different values of $\dot{M}_*$.

To interpolate a sink particle's $R_*$ and $T_{\mathrm{eff}}$ from tabulated tracks, instead of the sink's rapidly fluctuating instantaneous accretion rate, we used its lifetime averaged accretion rate
\begin{equation}
\langle \dot{M}_*\rangle (\leq t) = \frac{M_*(t)}{t-t_{\mathrm{form}}}\, , 
\end{equation}
where $t_{\rm form}$ is the sink formation time. The interpolated radii and temperatures were then used in Equations (\ref{eq:l_surf}) and (\ref{eq:l_acc}) to compute the sink's total luminosity. We neglected the luminosities of sinks with $M_*<0.1\,\msun$. This procedure produced stellar radii and effective temperatures in good agreement with the low-metallicity, disk-accretion models of \citet{Hosokawa10}. 


The geometry of the accretion flow can have a significant effect on protostellar 
evolution \citep[e.g.,][]{Hartmann97,Behrend01,Hosokawa09,Hosokawa10}. The model in which the accretion flow is spherically symmetric and in free fall is known as `hot accretion'. The material acquires a high entropy upon passing an accretion shock at the stellar surface and blankets the surface.  This produces bloated protostellar radii. If accretion is channeled through a circumstellar disk, the material loses entropy prior to joining the star and the blanketing effect is absent, so the protostellar radius is comparably smaller.  Accretion collimated 
by the protostellar magnetosphere may belong in between these limits, receiving an entropy increment in the accretion shock, but with little blanketing. 
Thus protostellar evolution is not uniquely determined by the accretion rate but also depends on the accretion geometry.  
By simulating the evolution of accreting protostars on the Hertzsprung-Russell diagram, \citet{Hosokawa11a} found that protostellar tracks with pure disk accretion significantly underpredict the luminosity of hot ($T>4000\,\kelvin$) protostars. 

Observationally, it is challenging to determine the accretion geometry due to the heavy obscuration of embedded protostellar sources. Disks are known to be common around massive stars \citep[e.g.,][]{Patel05,Cesaroni07} and simulations of massive star formation typically show circularization of accretion streams into disks \citep[e.g.,][]{Yorke02, Banerjee06,Krumholz07,Krumholz09}. These simulations, however, do not resolve the protostellar magnetosphere and cannot exclude the possibility that the material accreting through a disk eventually channels onto the magnetic poles. 

In \citet{SafranekShrader14a} we crudely modeled the accretion luminosity and neglected the intrinsic  luminosity. This simplification is valid for low-mass ($\lesssim3\,\msun$) protostars.  Specifically, we estimated the accretion luminosity with Equation (\ref{eq:l_acc}) after substituting the \citet{Stahler86} estimate of the protostellar radius
 \begin{equation}
R_* = 26R_{\odot} \left(\frac{M_*}{\msun}\right)^{0.27}\left(\frac{\dot{M}_*}{10^{-3}\,\msunperyr}\right)^{0.41}\,.
\label{eq:r_stahler}
\end{equation}
The latter estimate, however, assumes spherically symmetric, hot accretion. In contrast, we computed the \textsc{mesa} tracks consistent with the premise that gas arrives at the protostellar surface through a cold, thin disk. Since hot accretion models lead to protostellar radii potentially orders of magnitude larger than cold accretion models \citep[e.g.,][]{Hosokawa10},  the sink particle accretion luminosity in this present study is much larger than in \citet{SafranekShrader14a}.

Finally, we emphasize that our prescription for protostellar radiative feedback is entirely based on local approximations, modeling the effect of optical depth with the escape probability formalism.  Radiative transfer simulations like that of \citet{Krumholz12} are required to raise the modeling of primordial star formation to a new level of realism.



\label{fig:radial}

\renewcommand{\arraystretch}{0.4}
\begin{figure*}
 \begin{tabular}{l l | }
 \includegraphics[width=0.32\textwidth]{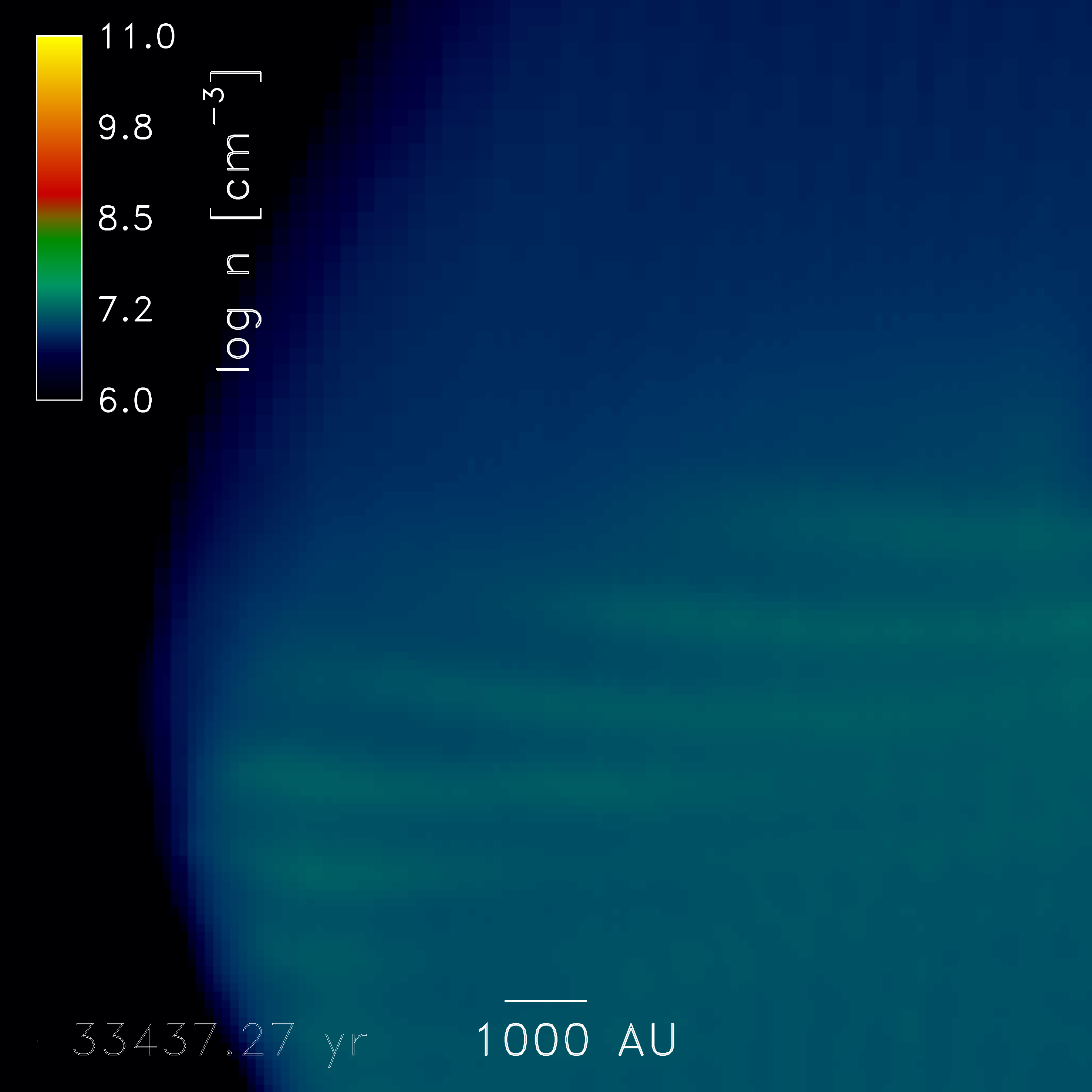} 
  \includegraphics[width=0.32\textwidth]{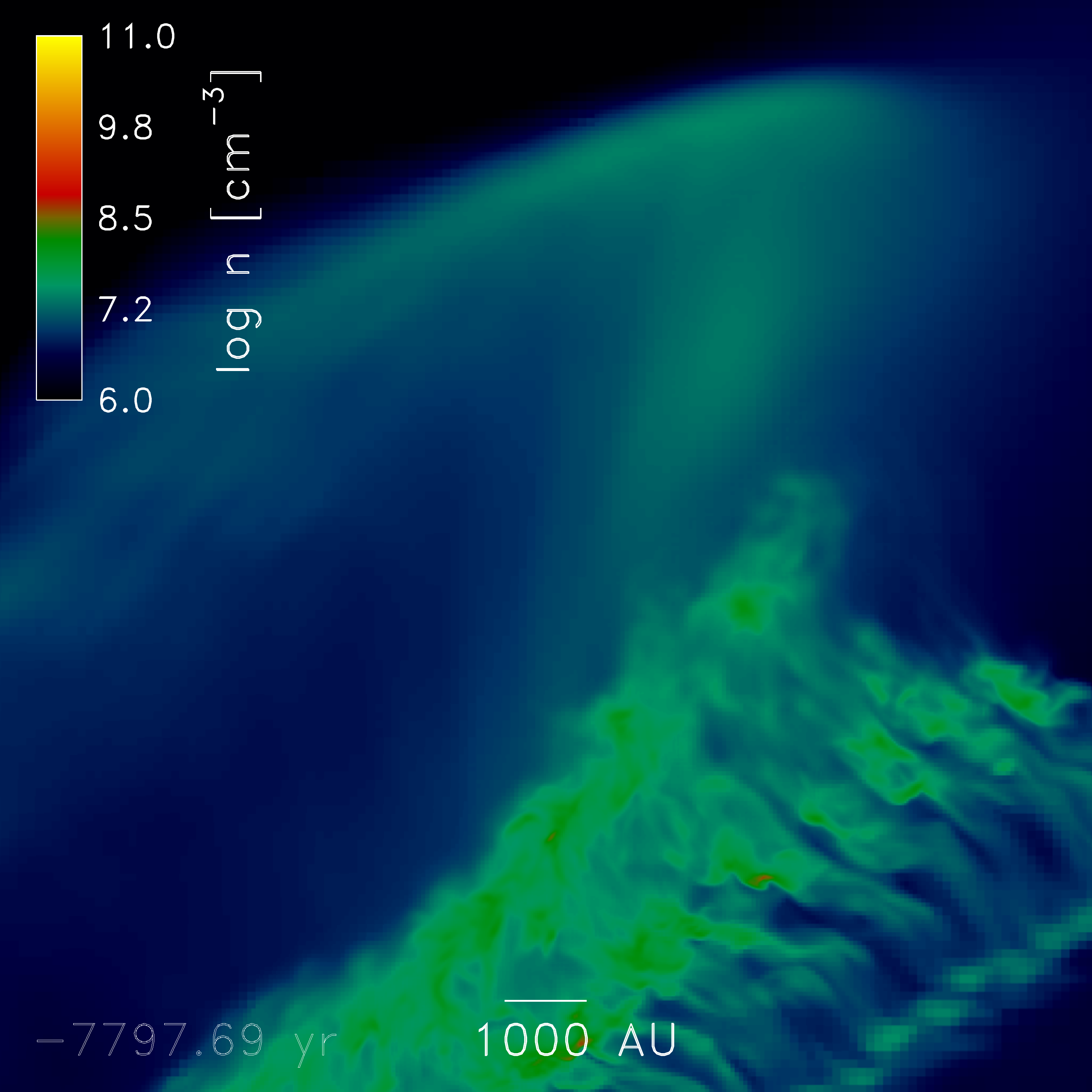}   
  \includegraphics[width=0.32\textwidth]{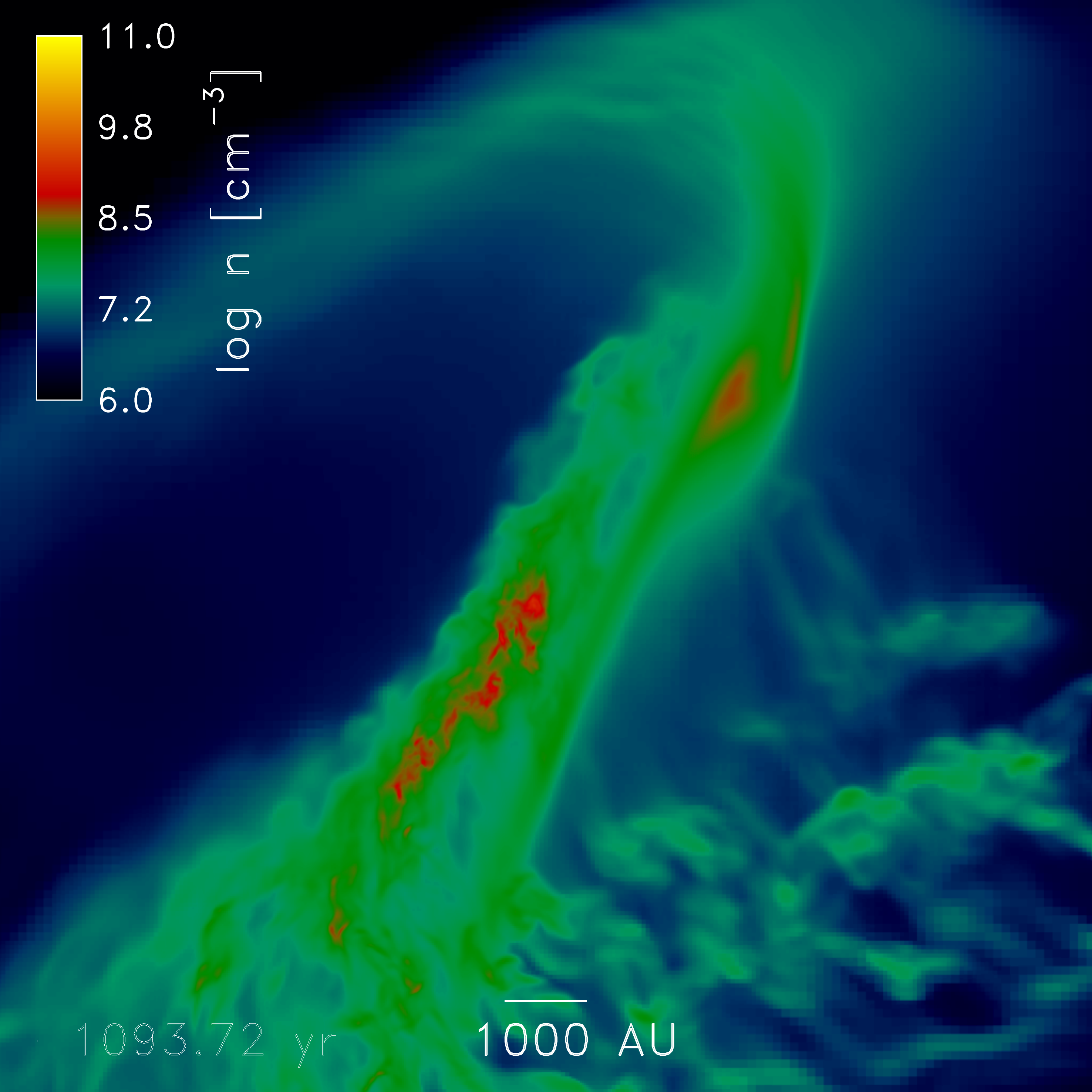}    \\
   \includegraphics[width=0.32\textwidth]{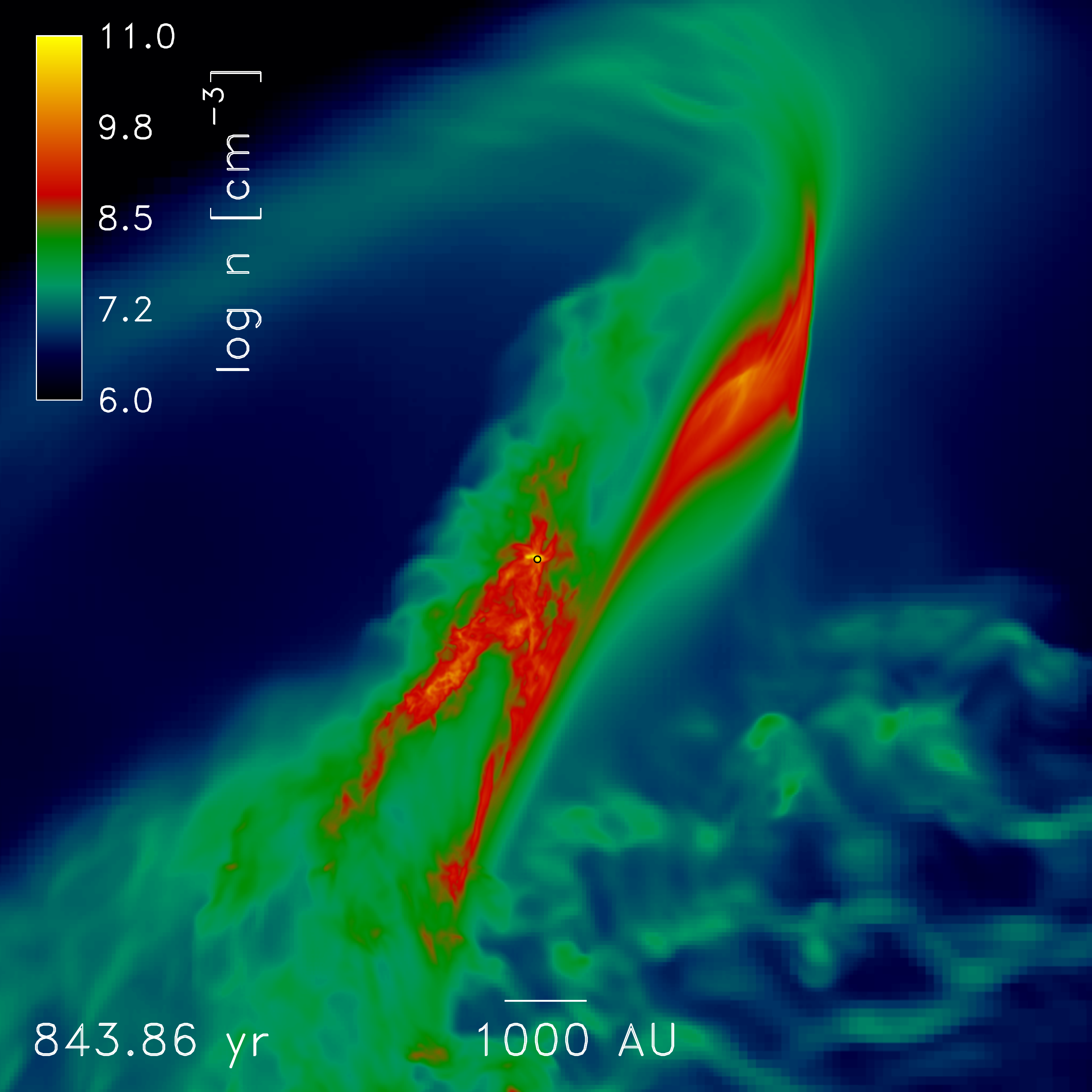} 
  \includegraphics[width=0.32\textwidth]{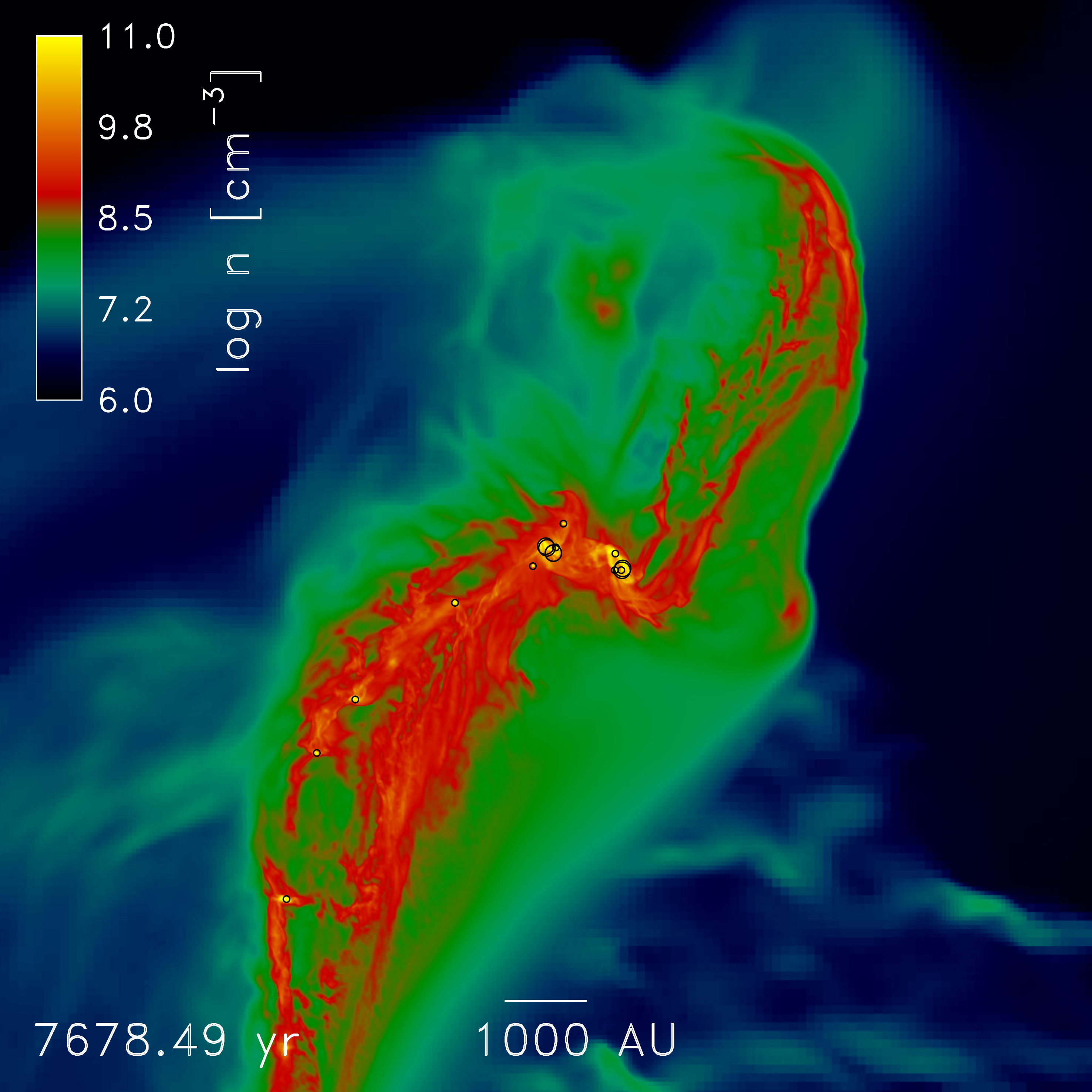} 
  \includegraphics[width=0.32\textwidth]{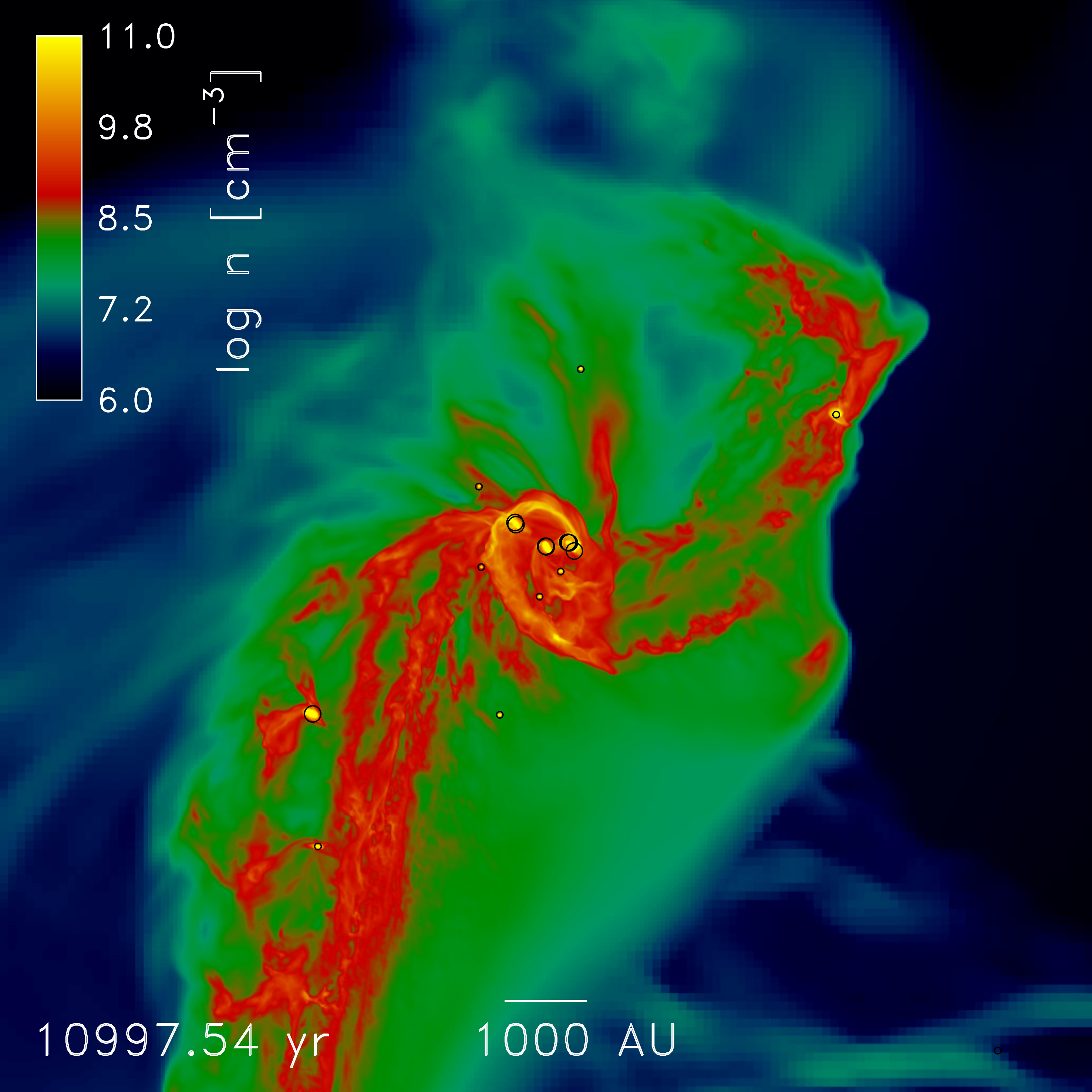}    \\
   \includegraphics[width=0.32\textwidth]{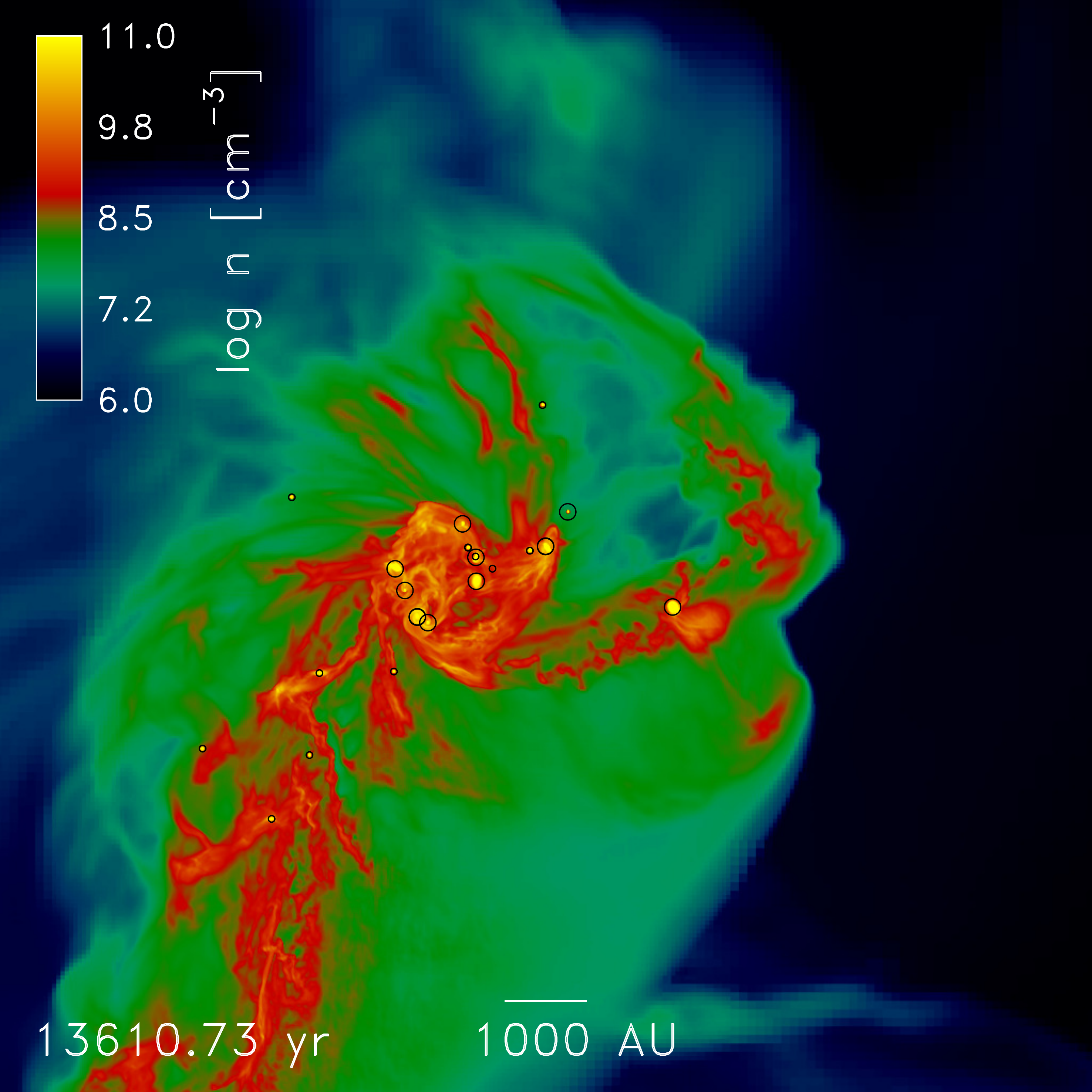} 
  \includegraphics[width=0.32\textwidth]{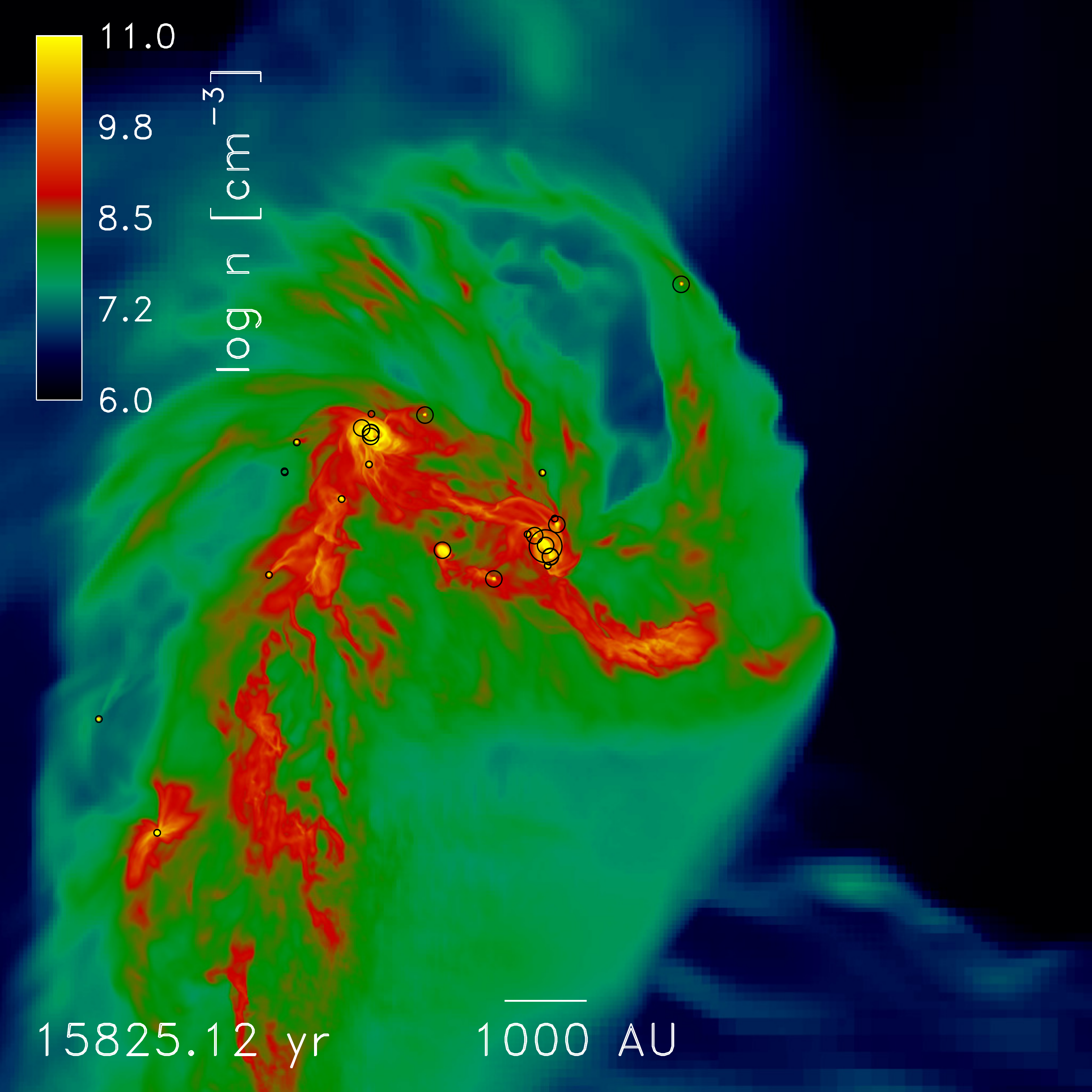} 
  \includegraphics[width=0.32\textwidth]{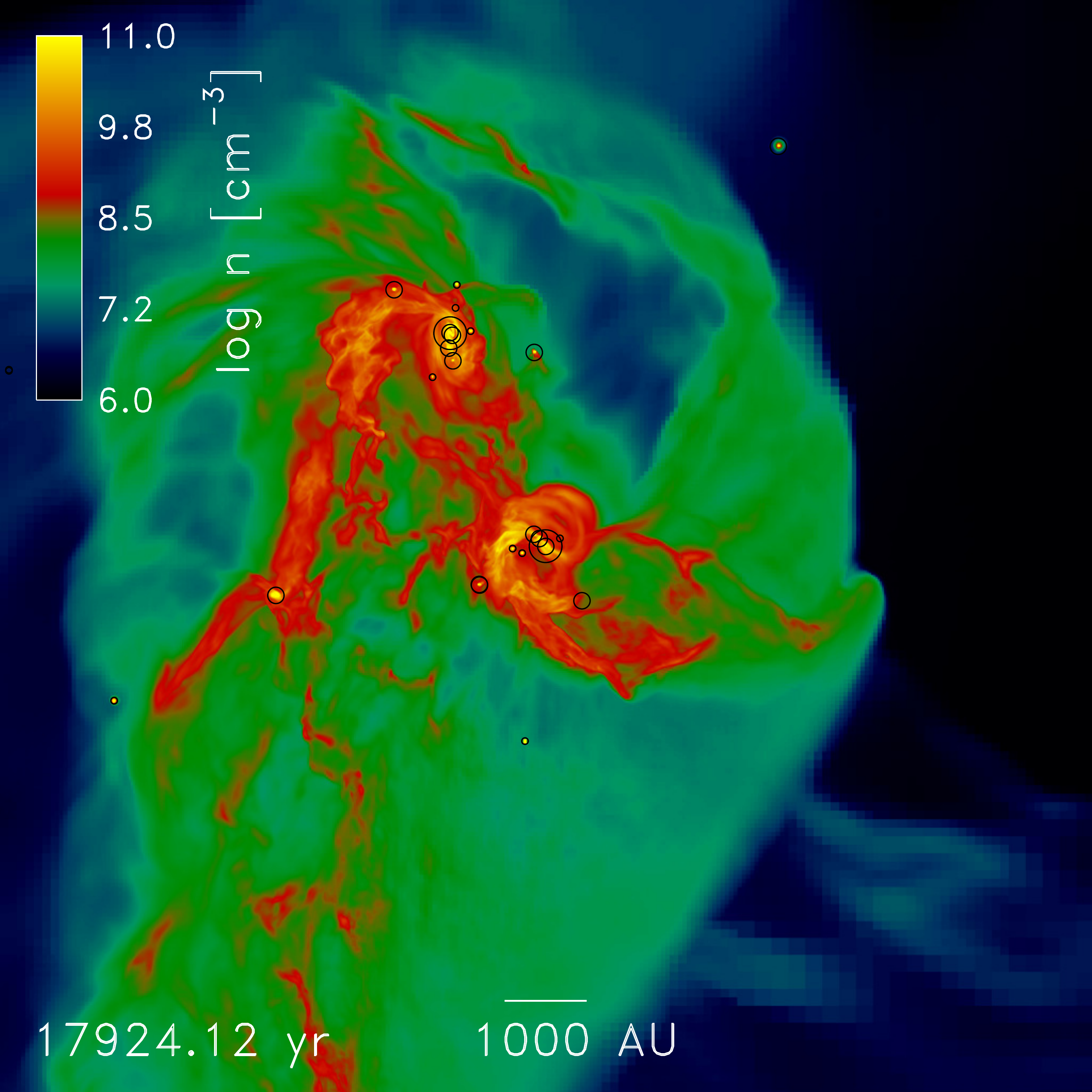}    \\

\end{tabular}
\caption{Mass-weighted line-of-sight projections of gas density in the $x-y$ plane. The top-left panel shows the initial state of the gas as extracted from the parent cosmological simulation. Time increases from left to right and from top to bottom, with $t=0$ shifted to the formation of the first sink. Circles mark projected sink locations with the circle size increasing with sink mass. The smallest sinks $M<\msun$ are drawn with twice the sink particle accretion radius $\racc=20\,\au$, the sinks with $1\,\msun<M<10\,\msun$ with $5\,\racc$, and the sinks with $M>10\,\msun$ with $10\,\racc$.}
\label{fig:dens_morph}
\end{figure*}

\renewcommand{\arraystretch}{0.4}
\begin{figure*}
 \begin{tabular}{l l | }
 \includegraphics[width=0.32\textwidth]{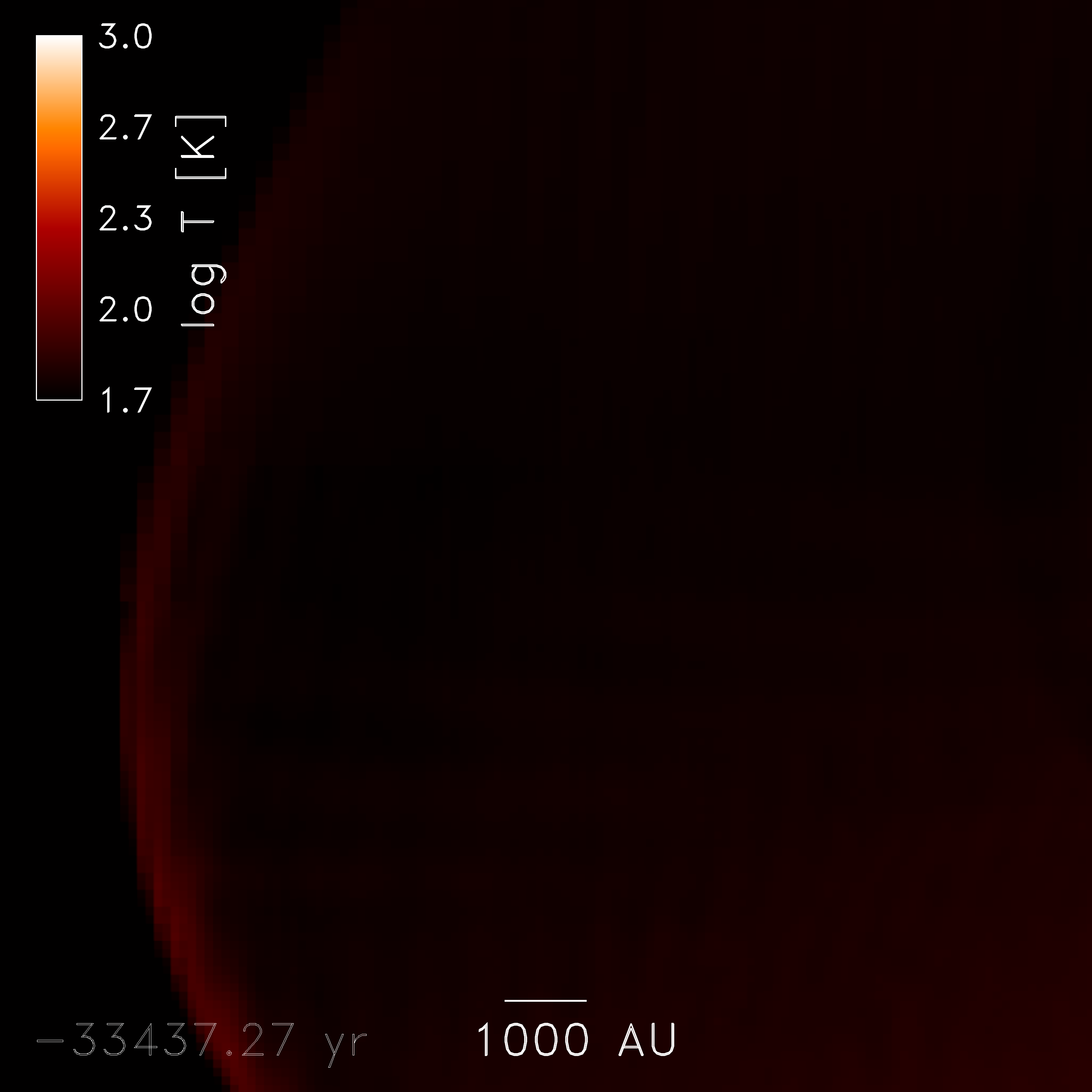} 
  \includegraphics[width=0.32\textwidth]{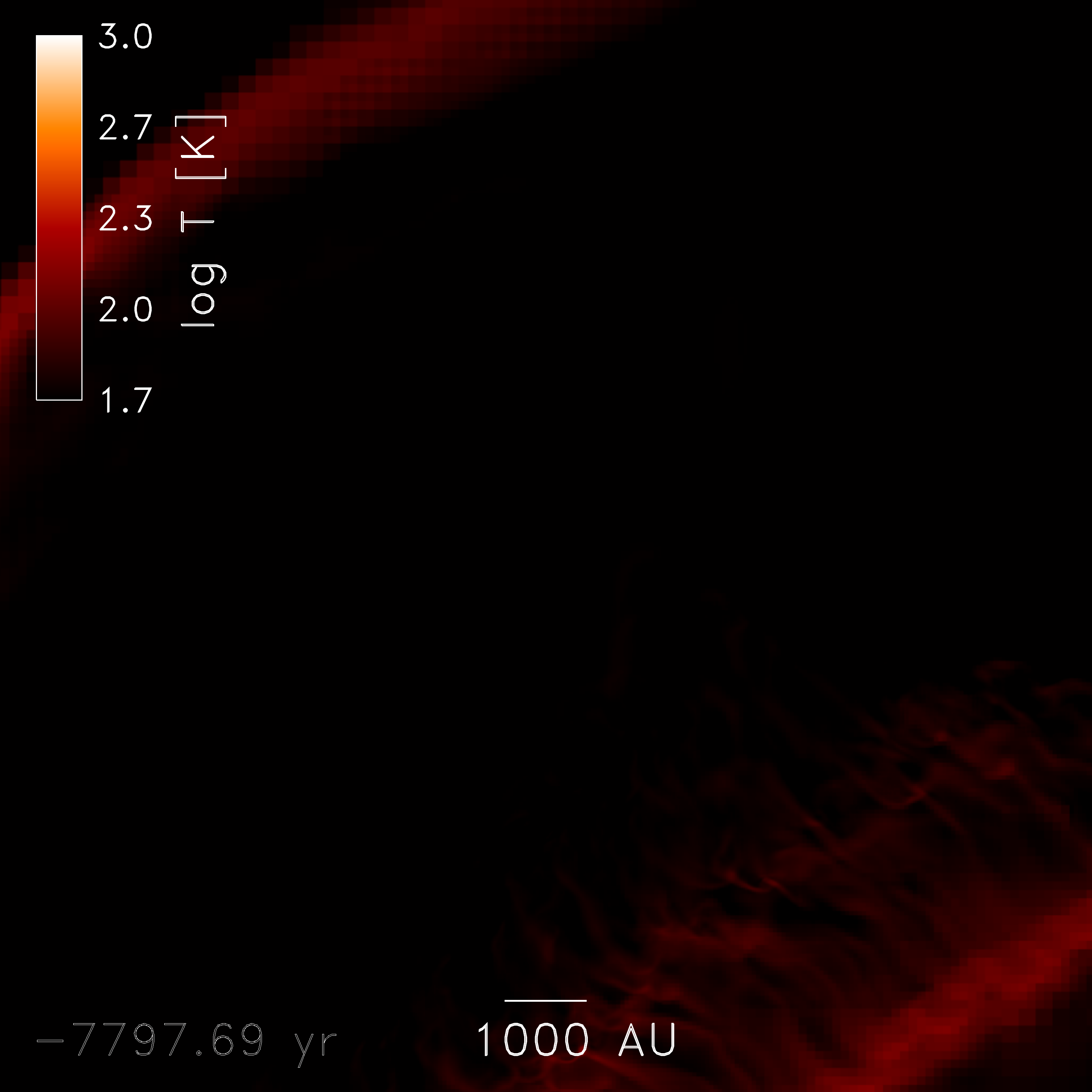} 
  \includegraphics[width=0.32\textwidth]{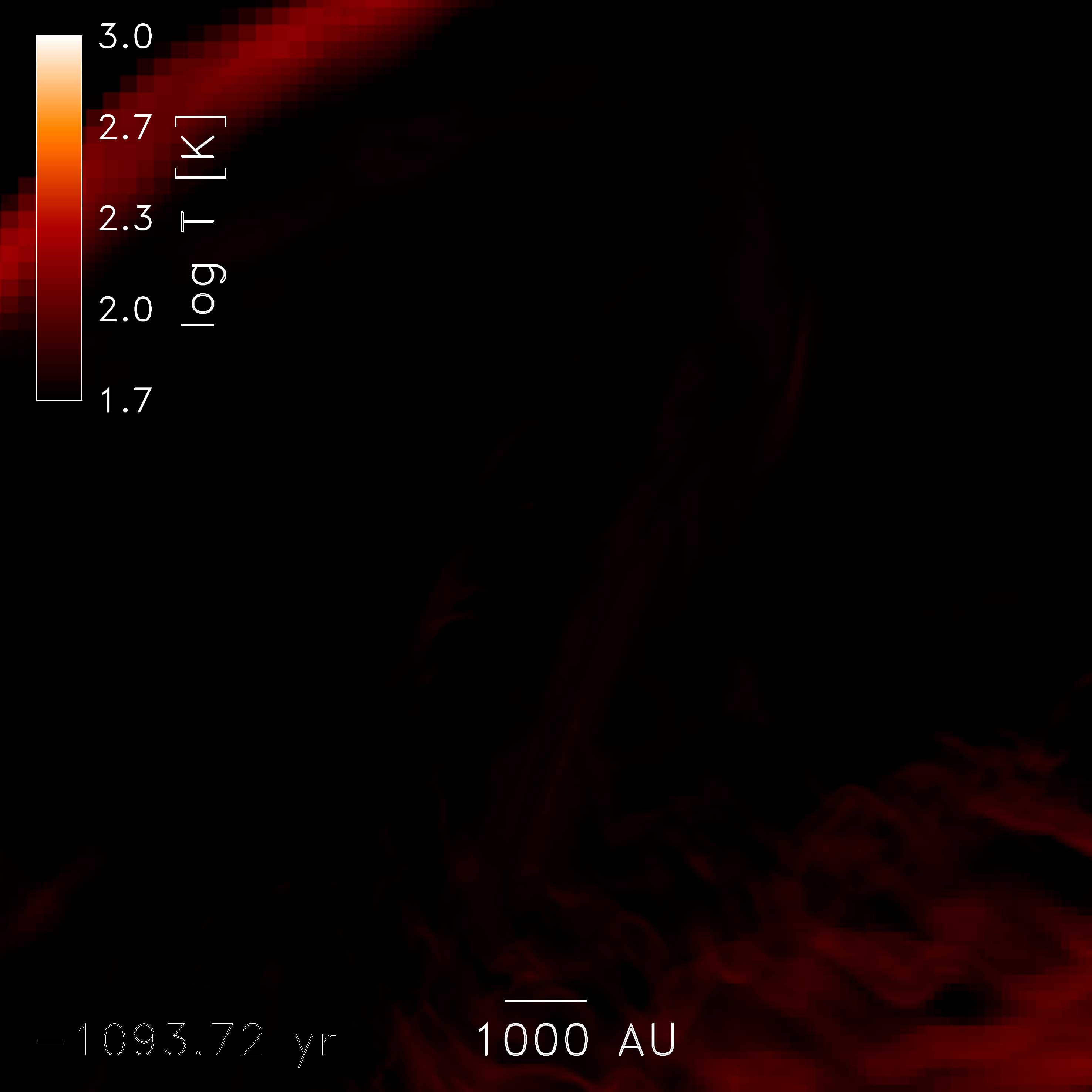}    \\
   \includegraphics[width=0.32\textwidth]{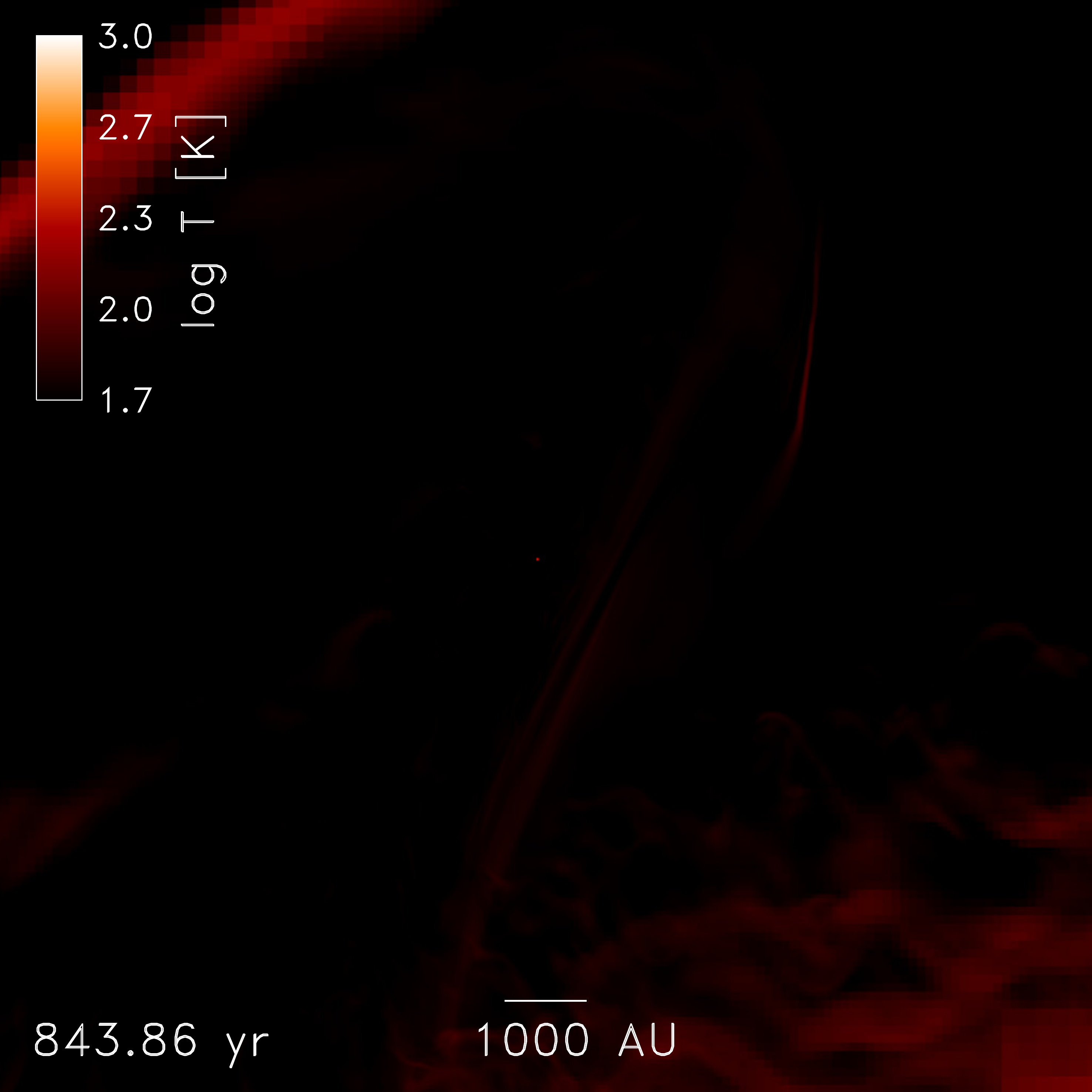} 
  \includegraphics[width=0.32\textwidth]{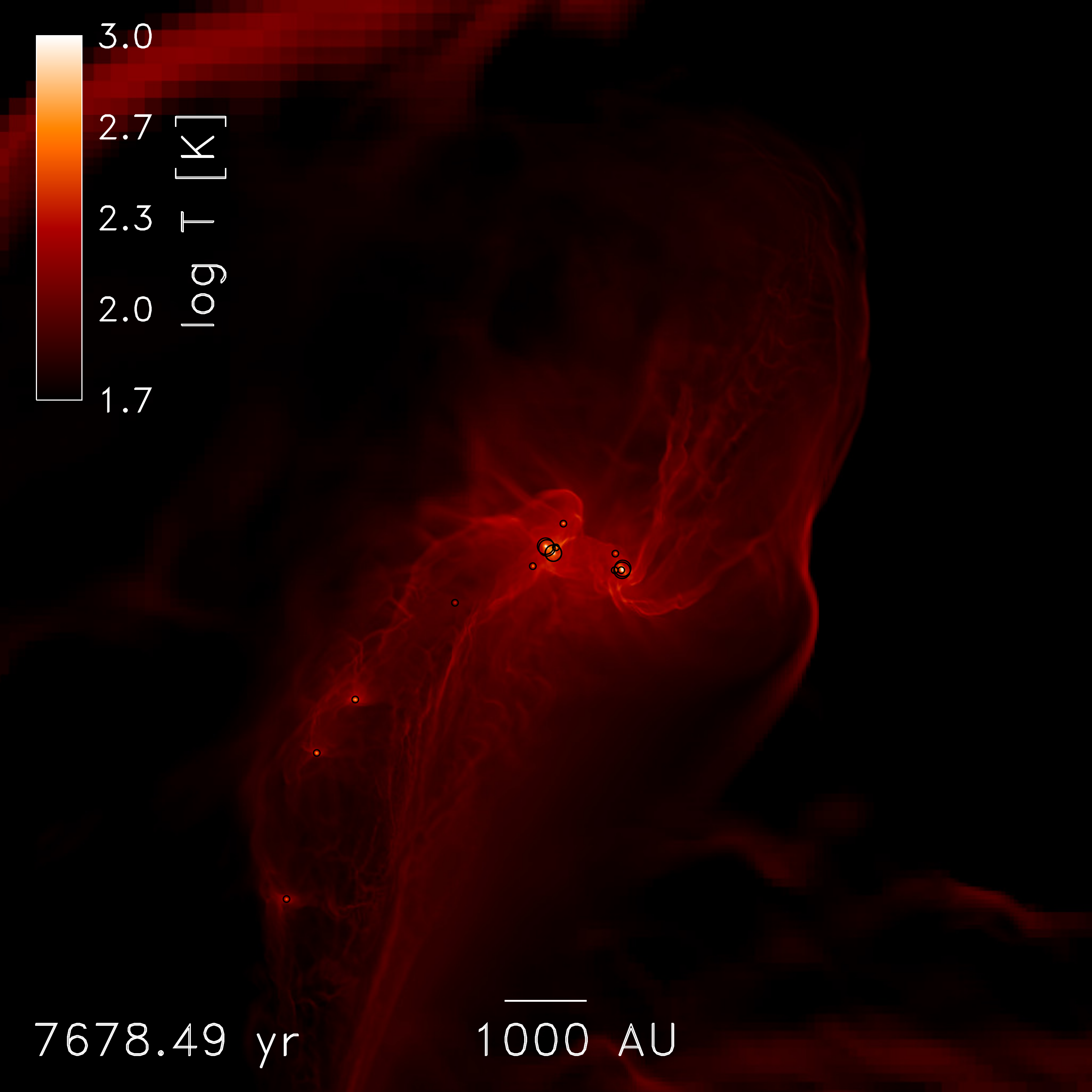} 
  \includegraphics[width=0.32\textwidth]{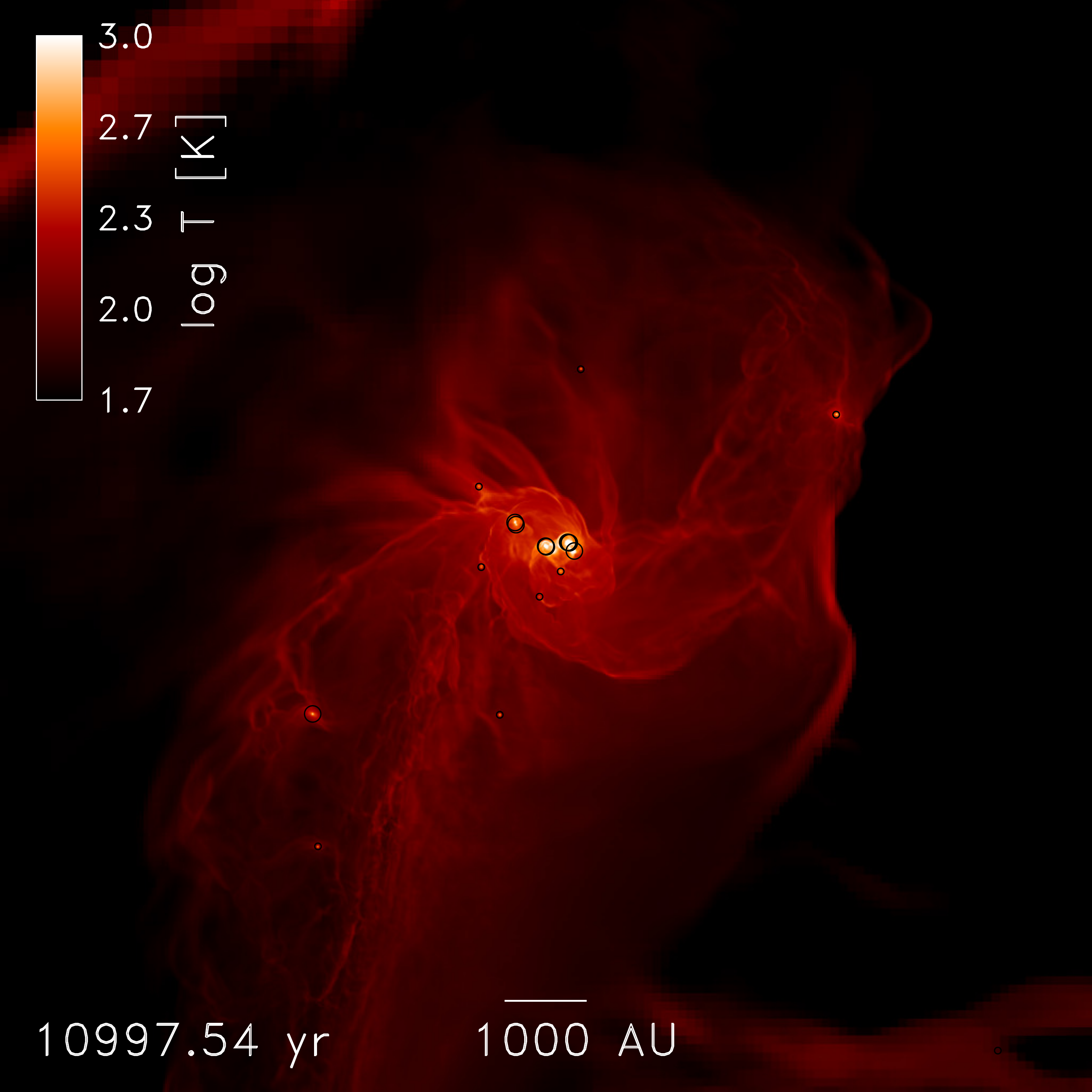}    \\
   \includegraphics[width=0.32\textwidth]{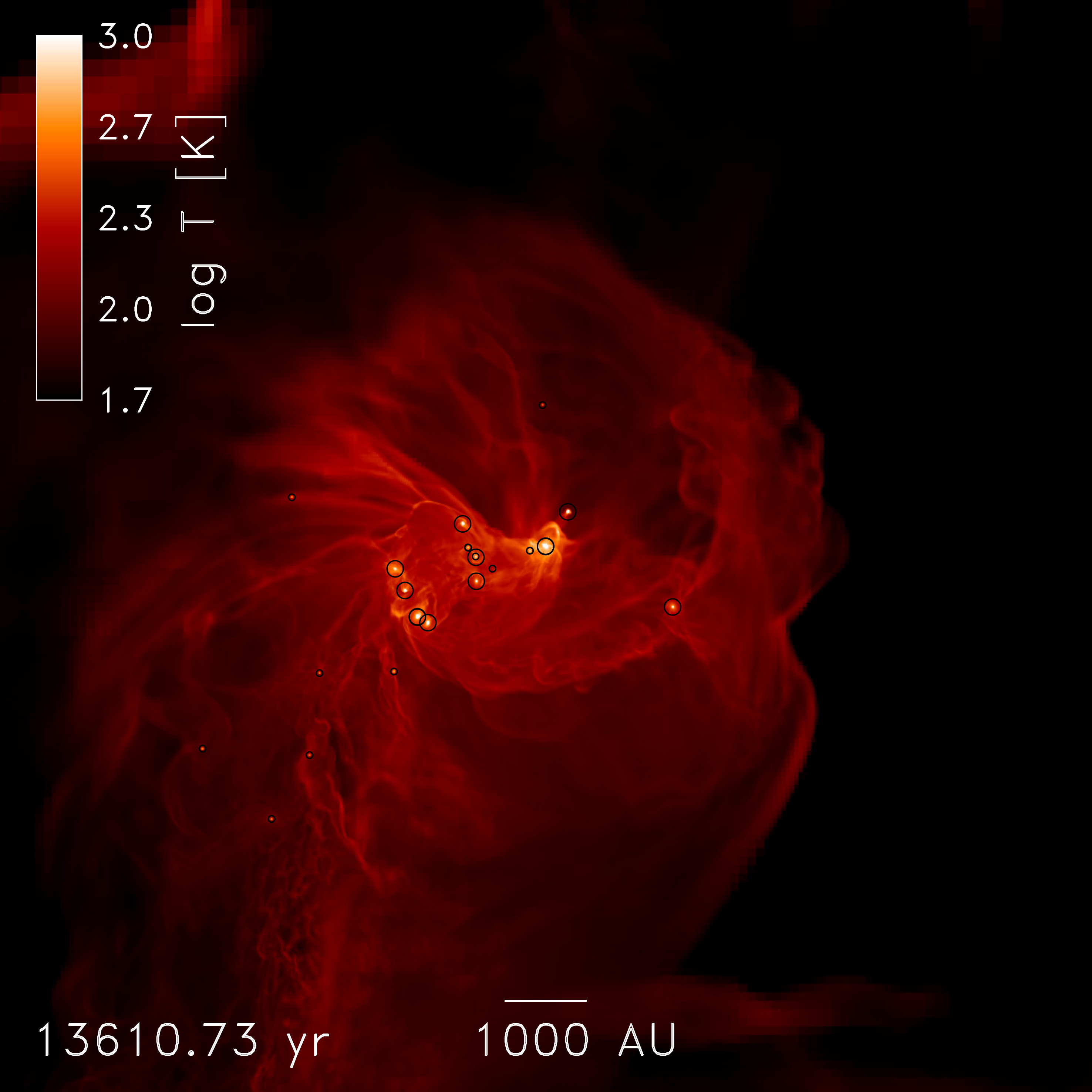} 
  \includegraphics[width=0.32\textwidth]{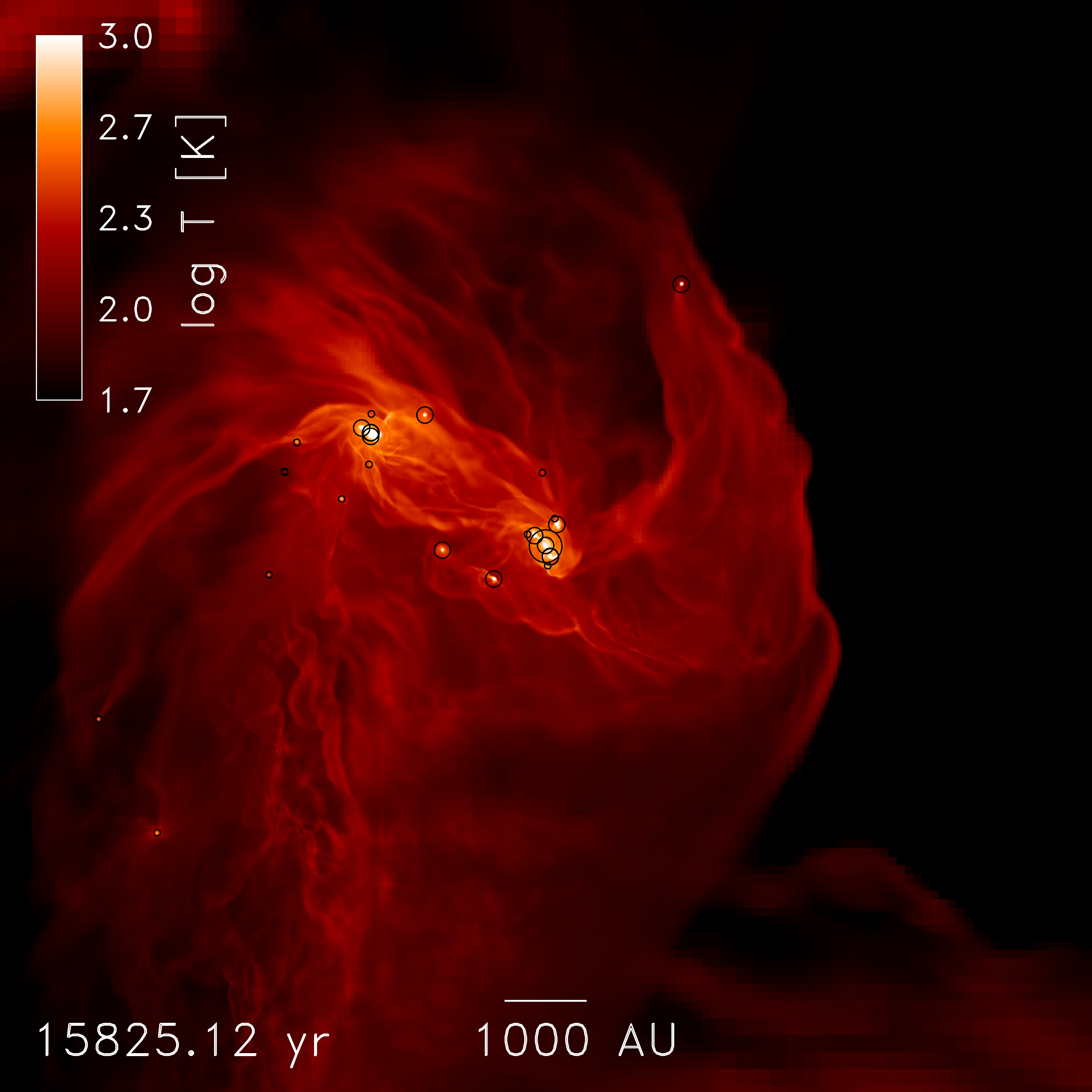} 
  \includegraphics[width=0.32\textwidth]{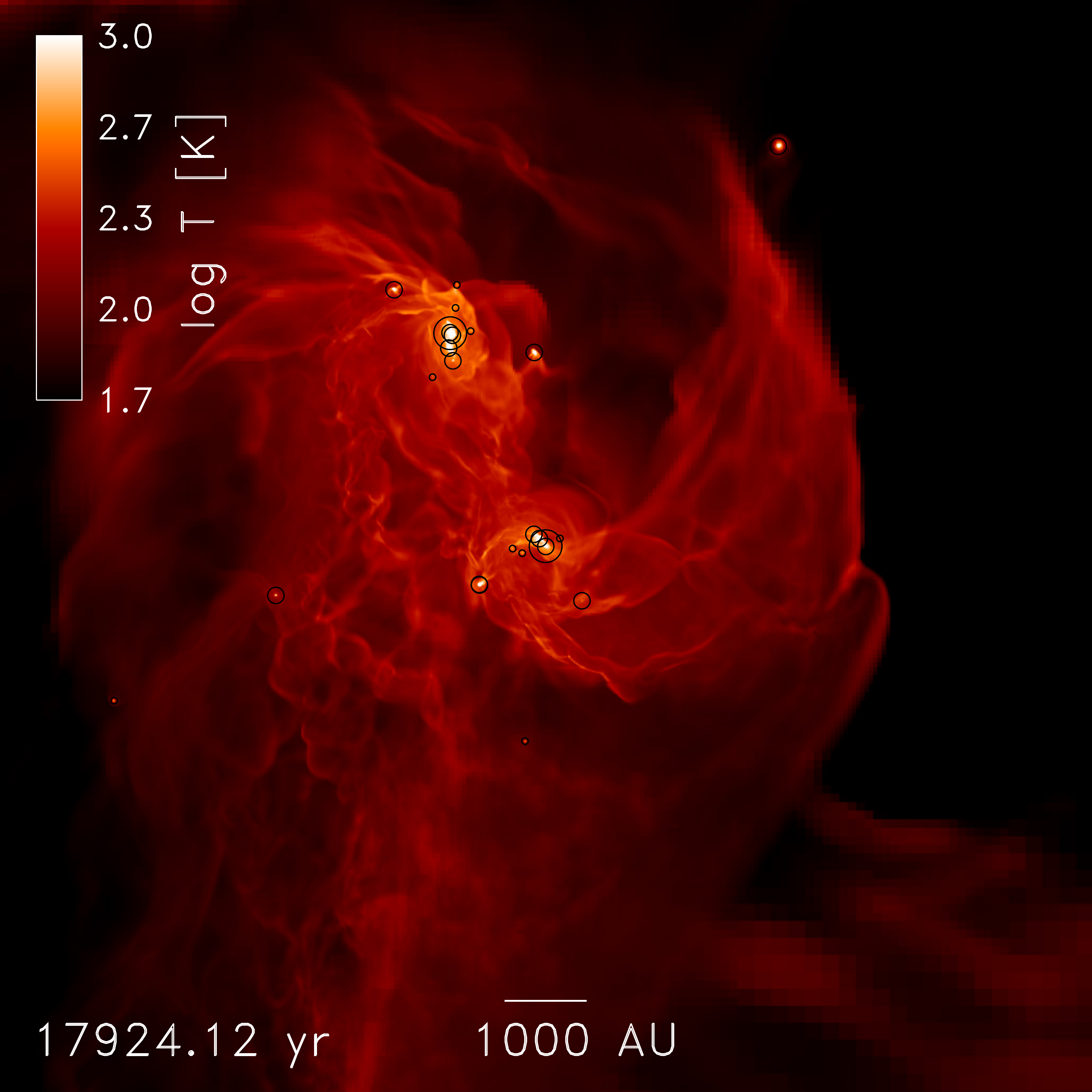}    \\

\end{tabular}
\caption{Same as Figure \ref{fig:dens_morph} but for gas temperature.}
\label{fig:temp_morph}
\end{figure*}

  \begin{figure*}
\includegraphics[width=0.95\textwidth]{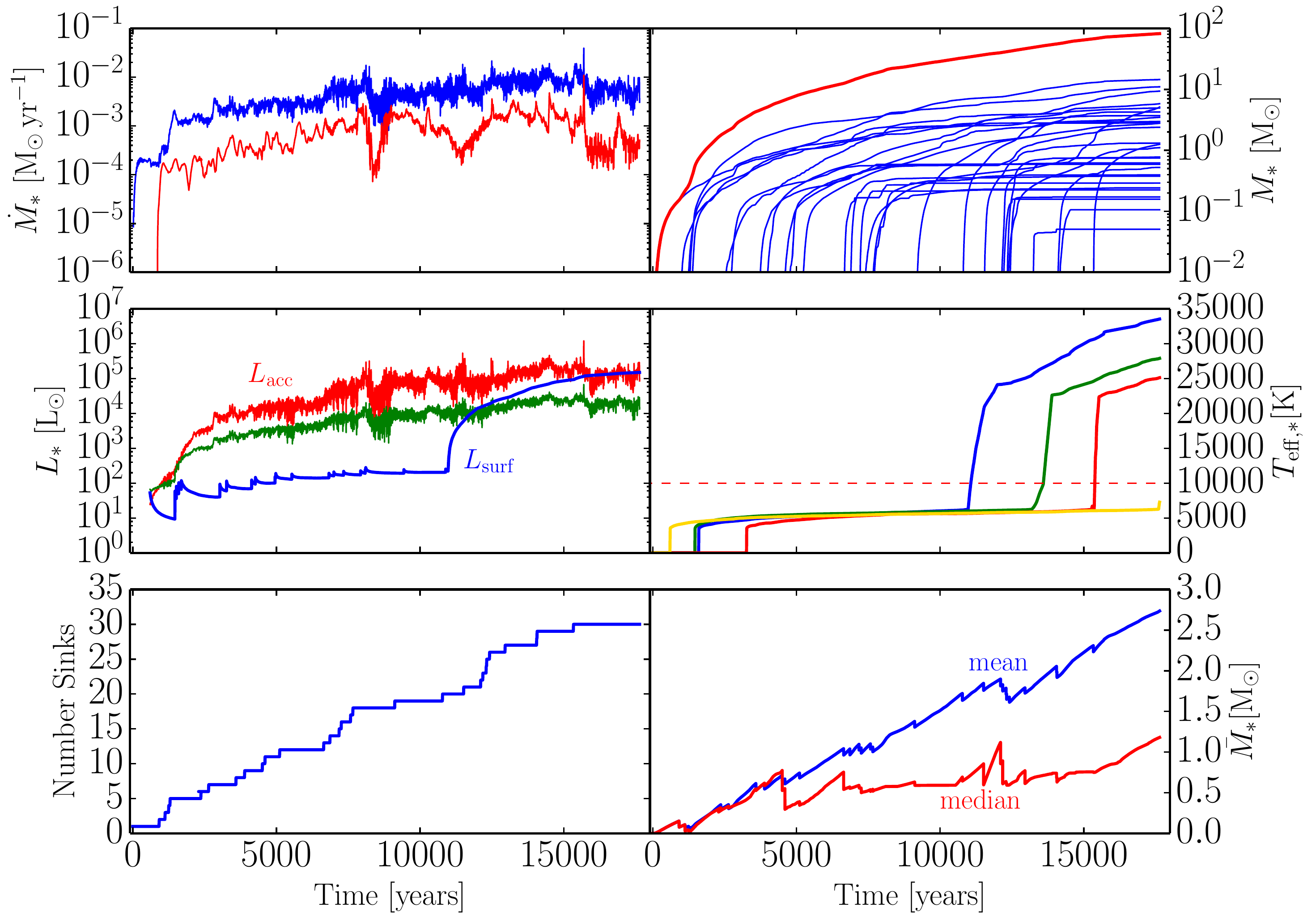}
\caption{Sink particle properties as a function time, with the time coordinate shifted so that the first sink forms at $t=0$. The top-left panel shows the total accretion rate onto sinks (blue) and the accretion rate onto the sink that grows the most massive by the end of the simulation (red). The top-right panel shows the total mass in sinks (red) and the masses of individual sinks (blue). The middle-left panel shows the total accretion luminosity (red) and total intrinsic luminosity (blue) for all sinks. It also shows, in green, what the total accretion luminosity would have been if the protostellar radii had been calculated using Equation (\ref{eq:r_stahler}) instead of with \textsc{mesa} evolutionary tracks. The middle-right panel shows the effective temperatures of four sinks that are the most massive at the end of the simulation. The dashed line at $T_{\rm eft}=10^4\,\kelvin$ is the approximate effective temperature limit above which a protostar would emit significant ionizing radiation. The bottom-left panel is the total number of sinks and the bottom-right panel shows the mean (blue) and median (red) sink mass.}
\label{fig:sink}
\end{figure*}


\section{Results}
\label{sec:results}


\subsection{Structure of the Collapsing Cloud}
\label{sec:morph}

The initial conditions, as extracted from the parent cosmological simulation, consist of an ellipsodial pre-stellar clump on the verge of gravitational collapse. The spherically averaged density profile is $\rho\propto r^{-1.5}$ from $\approx4\times10^4\,\au$ to the edge of the computational box. At radii $< 10^4\,\au$, the density profile is relatively shallower, and further inside $10^3\,\au$, it levels off to $\nh\approx10^7\,\cc$. Similar density profiles have been used by other groups to initialize isolated star-forming spherical clouds \citep[e.g.,][]{Krumholz12} and are consistent with observations of massive star forming clumps \citep[e.g.,][]{Beuther06}. 

Figure \ref{fig:dens_morph} shows snapshots of the morphological evolution of the central $\sim15,000$ AU from the initial ellipsoidal clump until $51\,\mathrm{kyr}$ later. The first sink forms $\approx 33\,\mathrm{kyr}$ after the beginning of the simulation in gas compressed by a large-scale flow convergence inherited from the parent cosmological simulation. In the next $\sim 5\,\mathrm{kyr}$, the bulk of sink formation occurs in a pair of turbulent, rotating, disky structures separated by $\sim 2000$ AU, as seen in the middle-left panel of Figure \ref{fig:dens_morph}. The structures subsequently merge into a larger such structure.  Still later, a new pair of rotating disk structures is apparent.  Additional sink formation occurs in two linear filamentary features extending in opposite directions, as seen in the middle panel of Figure \ref{fig:dens_morph}.  Close encounters between sinks are common.  Numerous sink ejections occur as a result of close encounters and many-body interactions. The sink-gas system reaches approximate virial equilibrium relatively quickly, in $\sim5\,\mathrm{kyr}$, after the onset of sink formation.

Figure \ref{fig:temp_morph} shows the evolution of the mass-weighted gas temperature and is otherwise identical to Figure \ref{fig:dens_morph}. Within $\sim8\,\mathrm{kyr}$ after the onset of sink formation, the dust temperature in the central $\sim2000\,\au$ begins to exceed $\sim100\,\kelvin$ and the dust starts heating the gas. At the end of the simulation, the central $\sim3000\,\au$ has an average gas temperature $\sim300\,\kelvin$, hot enough to significantly raise the thermal Jeans mass and inhibit new sink formation.

  \begin{figure*}
\includegraphics[width=0.95\textwidth]{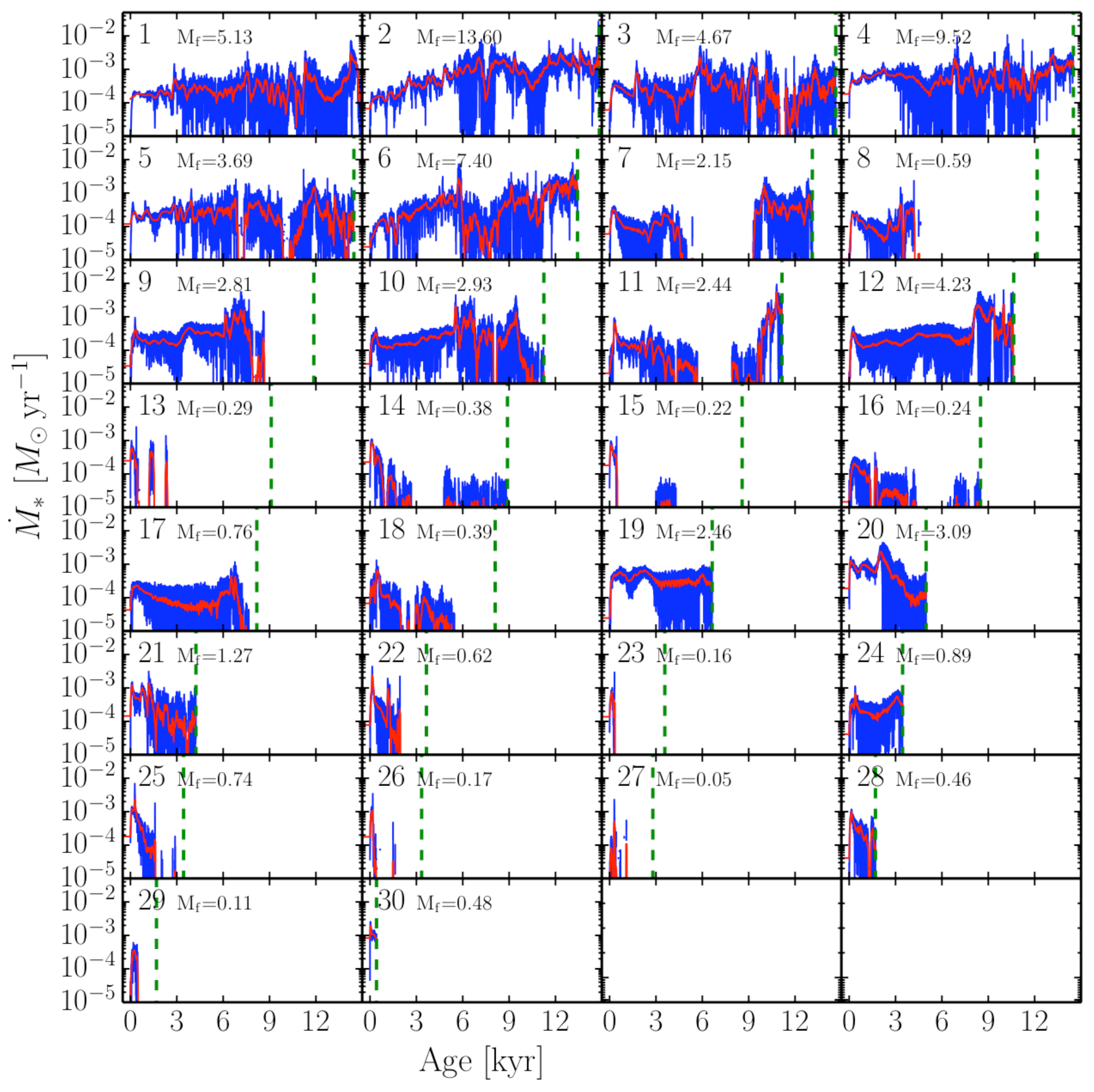}
\caption{Instantaneous accretion rates of individual sink particles as a function of time since the individual sink's birth. The birth rank and the final mass at the end of the simulation $M_{\rm f}$ are shown in the top-left corner of each panel. The mass of the sink at the final simulation snapshot, , is also shown for each sink. The blue line shows the accretion rate extracted directly from the simulation and the red line is the accretion rate binned over a $300$ year period. The vertical dashed green line in each panel shows the end of the simulation. Panels in which the accretion rate drops below $10^{-5}\,\msunperyr$ (e.g., panels 8, 13, 15, and 23) show sinks that have experienced suppression or termination of accretion, usually after a close encounter with more massive sinks.}
\label{fig:sink_accretion}
\end{figure*}

\subsection{Sink Particle Formation and Growth}
\label{sec:cluster}

We ran the simulation for $18\,\mathrm{kyr}$ after the formation of the first sink. In this period, $30$ sinks formed with a total mass of $81\,\msun$. For most of this time the median sink mass stayed constant at $\approx 0.5\,\msun$, though in the last $\sim3\,\mathrm{kyr}$, the median mass increased above $1\,\msun$. At the end of the simulation, the most massive sink was $\approx14\,\msun$ and was undergoing Kelvin-Helmholtz contraction towards the main sequence.

Figure \ref{fig:sink} displays a selection of sink properties as a function of time: the total accretion rate onto all sinks $\dot{M}_{*,\mathrm{tot}}$, the total accretion luminosity and intrinsic luminosity from all sinks, the total number of sinks, the total mass in sinks and the masses of individual sink particles, the effective temperature of the four sinks that would grow to be the most massive ones by the end of the simulation, and the mean and median sink particle mass.

Beginning $\approx 2\,\mathrm{kyr}$  after the formation of the first sink and until the end of the simulation, the total mass accretion rate onto all sinks grew from $\approx10^{-3}\,\msunperyr$ to almost $10^{-2}\,\msunperyr$. In the final $2\,\mathrm{kyr}$, there is indication of a slight downturn in $\dot{M}_{*,\mathrm{tot}}$, though it is unclear if this is a genuine trend or a transient accretion rate fluctuation. However, there is indeed a significant paucity of new sink formation in the final $\approx4\,\mathrm{kyr}$. 


In Figure \ref{fig:sink_accretion} we show the accretion rate as a function of time since birth for each of the $30$ sinks that form in the simulation. These accretion histories vary widely, but some general conclusions can be drawn. With very few exceptions and independent of the final sink mass, each sink accretes very slowly at birth, $\dot{M}_*<10^{-6}\,\msunperyr$. The accretion rate rises to between $10^{-4}\,\msunperyr$ and $10^{-3}\,\msunperyr$ in less than $1\,\mathrm{kyr}$. We refer to this rise as the initial accretion rate peak. Such accretion rates are $\sim2-20$ times the nominal core accretion rate $\sim \cs^3/G$ expected in a marginally Jeans-unstable collapsing core with a temperature $T\approx50\,\kelvin$ \citep[e.g.,][]{Shu77}. 

The instantaneous accretion rate onto a given sink also exhibits strong temporal fluctuations, often by more than three orders of magnitude on time scales $\ll1\,\mathrm{kyr}$. Some of this variability is unphysical numerical discreteness of our operator-split scheme for accreting gas mass onto  sinks. When $\dot{M}_*$ is binned on $300$ year time scales (red lines in Figure \ref{fig:sink_accretion}), short term variation is still present but with a lower amplitude. This variability in the accretion rate translates directly into variable accretion luminosity via Equation (\ref{eq:l_acc}). Observed embedded protostars typically have luminosities one to two orders of magnitude smaller than what would be naively expected from indirect estimates of accretion rates \citep[e.g.,][]{Kenyon94,Evans09,Enoch09}. Highly variable, or episodic, accretion is a natural consequence of protostellar growth in a turbulent environment or from gravitationally unstable circumstellar disks.  Variable accretion is one potential solution to the `protostellar luminosity problem' \citep[e.g.,][]{Dunham12}.

Sinks that ultimately reach high mass, here arbitrarily defined as $M_*>3\,\msun$ (e.g., panels 1, 2, 3, 4, 5, 6, and 12 in Figure \ref{fig:sink_accretion}), do so via extended periods of rapid accretion, with $\dot{M}_*\gtrsim10^{-4}\,\msun\,\mathrm{yr}^{-1}$ over $\gtrsim10\,\mathrm{kyr}$. Most of this accretion occurs in competition with other sinks and does not sample the material originally exclusively bound to, or even geometrically associated with the sink. The lower panels of Figure \ref{fig:dens_morph} show that a sink's initial Jeans-unstable progenitor core is typically accreted entirely within $\sim3\,\mathrm{kyr}$. At the end of the simulation, many massive sinks sustain high accretion rates, although we caution that our simulation does not include two principal mechanisms that could suppress accretion, protoionization and protostellar outflows.  A modeling of protostellar outflows \citep[e.g.,][]{Cunningham11,Federrath14}, photoionization \citep[e.g.,][]{Peters10,Dale11}, and a more accurate treatments of radiative transfer, would increase the physical realism, particularly with regards to the formation of massive stars and the driving of interstellar turbulence.

Low mass sinks with $M_*<1\,\msun$ (e.g., panels 8, 13, 14, 15, 18, 22, 23, 26, and 27 in Figure \ref{fig:sink_accretion}) typically exhibit sudden accretion rate termination, with the rate dropping below $10^{-5}\,\msunperyr$ and in some cases much lower. This can be attributed to the relatively compact star forming environment ($\sim10^4\,\au$) in which close encounters between sinks are commonplace. Many-body encounters give large kinetic energy kicks to the lowest-mass bodies, placing them on fast or even unbound orbits.  Such orbits sample lower density gas, which in turn reduces the Bondi-Hoyle accretion rate ($\dot{M}_{\rm BH}\propto \rho v^{-3}$).  Many-body encounters and ejections of this type become progressively common as the crowding of the star forming complex increases.

We observe an interesting correlation between birth rank and the initial peak accretion rate sinks achieve in their first $\sim1\,\mathrm{kyr}$. The earliest sinks (panels $1-12$) have typical initial peak accretion rates $\sim3\times10^{-4}\,\msunperyr$ while later sinks (panels $17-30$) have initial peak rates typically exceeding $10^{-3}\,\msunperyr$. This can be understood as a consequence of stellar radiative feedback from existing protostellar sources.  Protostars heat the central star forming environment to an increasing degree and this raises the core accretion rate $\sim \cs^3/G\propto T^{3/2}$ in newly collapsing cores (see Figure \ref{fig:temp_morph}). Many of these late-forming sinks with high initial peak accretion rates terminate accretion shortly after birth (e.g., panels 22, 23, 25, 26, and 27 in Figure \ref{fig:sink_accretion}) thanks to close encounters with massive sinks that are common in the dense star-forming complex.

We observe evidence for competitive accretion between the protostars in which accretion of initially unbound gas from a common gas reservoir determines the final protostellar mass \citep[e.g.,][]{Bonnell98,Bonnell01,Bate05a}. More massive stars not only have larger accretion cross sections, but they also segregate toward the cluster center, where the gas density, and thus accretion rate, are higher.  Elementary analytical models of this `rich-get-richer' scenario produce self-similar protostellar growth and predict a scale-free, power-law mass distribution akin to the observed high-mass stellar IMF \citep[e.g.,][]{Bonnell01a}.  Nevertheless, a predictive, realistic theory of the IMF taking into account competitive accretion remains lacking.  Models \emph{not} allowing for the dynamically complex competitive accretion, but rather assuming that stellar masses are imprinted by turbulent density fluctuations preceding star formation, have had more success in producing definitive predictions for the IMF \citep{McKee02,McKee03,Krumholz05a,Padoan02,Hennebelle08,Hopkins12}, in spite of the inevitable role of competitive accretion.  
Our sinks continue growing well after accreting the initial gravitationally bound `core' 
(Figure \ref{fig:sink_accretion}) by accreting in a relatively crowded star-forming environment (Figure \ref{fig:dens_morph}).  The sinks most massive at the end of the simulation acquire the bulk of their mass in the latter regime. This underscores the importance of the competitive-accretion-like mode of gas accretion, at least in the specific regime explored here.

We should note that the initial conditions play a crucial role for the mode of gas accretion and stellar growth, as has been emphasized by numerous authors \citep[e.g.,][]{Krumholz05a,Bonnell06a,Girichidis11}. In particular, simulations initialized with decaying turbulence tend to produce a coherent, central collapse, ideal conditions for competitive accretion. Those initialized from driven turbulence, in which small scale turbulence is continually replenished from larger scales, seem instead to fix stellar masses through fragmentation of turbulent structures into self-gravitating cores. The initial conditions of the present simulation, which were extracted from a parent cosmological simulation at a site of thermal instability, can be regarded as belonging in the former category, containing turbulence initially excited by thermal instability and then decaying, or growing if amplified by gravitational compression \citep[e.g.,][]{Robertson12}.

  \begin{figure}
\includegraphics[width=0.47\textwidth]{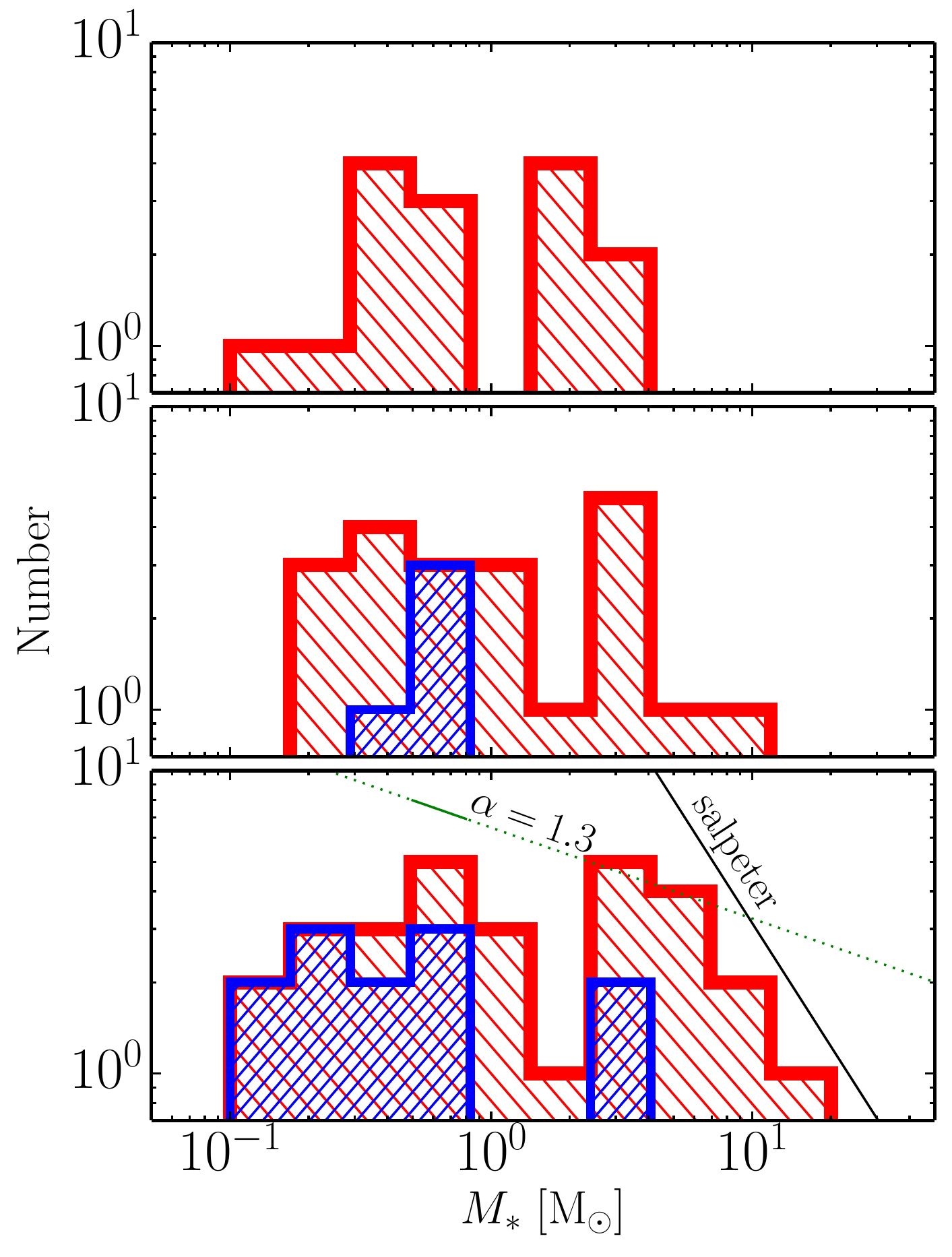}
\caption{Sink particle mass function at $7.6$ kyr (top panel), $11$ kyr (middle panel), and $18$ kyr (bottom panel) after the onset of sink formation. The red histograms are for all sink particles and the blue histograms are only for non-accreting sinks, namely those with instantaneous accretion rates below $10^{-7}\,\msunperyr$ in the last $500\,\mathrm{yr}$ of the simulation. The black and green straight lines indicate pawer-low slopes of $\alpha=2.35$ \citep{Salpeter55} and $\alpha=1.3$ \citep{Geha13} where $dN/dM_* \propto M_*^{-\alpha}$. The solid portion of the Geha et al.\ line, between $0.5\,M_\odot$ and $0.8\,\msun$, indicates the narrow stellar mass range in which observations constrain the IMF in two ultra-faint dwarf spheroidal satellites.}
\label{fig:sink_mf}
\end{figure}

\subsection{Sink Particle Mass Function}
\label{sec:imf}

In Figure \ref{fig:sink_mf} we show the mass function of sink particles between $0.1\,\msun$ and $20\,\msun$ in three representative simulation snapshots. The mass function evolves considerably over $\sim10$ kyr, particularly in view of the late formation of $>3\,\msun$ sinks. In blue, we also show the mass function of sinks that have `stopped growing', here defined as the sinks with accretion rates remaining below $10^{-7}\,\msunperyr$ for the final $0.5\,\mathrm{kyr}$ of the simulation. These are typically the sinks that have had a strong dynamical encounter with higher mass sinks.
All but two of the non-accreting sinks have masses below $1\,\msun$ at the end of the simulation.

Above $\sim 3\,\msun$, the sink mass function develops hints of a power-law-like tail toward higher masses. Since there are only $12$ sinks with such masses, the statistics is insufficient to perform estimation of the power-law slope.  We also caution that the sink mass function has \emph{not} converged at high masses. Had we run the simulation longer, the high-mass tail would have undoubtedly grown, tempered, perhaps, by the effects of photoionization that would be essential to include in the high-mass regime.

In the early-results version of the present simulation \citep{SafranekShrader14a}, which had not been run long enough to form high mass $\gtrsim 3\,\msun$ sinks, we tentatively attempted to constrain the power-law slope of the sink mass function in the intermediate mass range $0.1\,\msun< M_* < 3\,\msun$.  We obtained a maximum likelihood slope of $\alpha\approx1.3$ such that $dN/dM_* \propto M_*^{-\alpha}$.  Perhaps coincidentally, this slope was consistent with the stellar mass function slope $\alpha=1.3$ measured in the narrow stellar mass range $0.5\,\msun\lesssim M_*\lesssim 0.8\,\msun$ in ultra-faint dwarf (UFD) satellite galaxies Leo IV and Hercules \citep{Geha13}.  In the present, extended simulation, the mass function in the range $0.1\,\msun< M_* < 3\,\msun$ seems flat with $\alpha\approx 1$, similar to, and perhaps even shallower than reported in \citet{Geha13} and \citet{SafranekShrader14a}, albeit with  bin-to-bin statistical fluctuations, including an apparent, but statistically insignificant trough at $M_*\sim 2\,\msun$. 

We reiterate, however, that all sinks above $1\,\msun$ are still accreting at the end of the simulation (Fig.~\ref{fig:sink_mf}) and the sink mass function is far from convergence.  The results reported here should be construed as diagnostic of a trend in the protostellar mass function assembly process, rather than predictive of the final stellar IMF.  It is also important to emphasize that the sink particle mass need not directly translate in the final stellar mass because only a fraction of the mass accreted by a sink particle is incorporated into the final star. For example, mass loss to protostellar outflows is expected reduce the core-to-star conversion efficiency by $\approx50\%$ \citep{Matzner00,Alves07,Enoch08}.

There are other effects that could suppress gas accretion and inhibit further evolution of the mass function. H\,II regions produced by massive stars disperse natal molecular clouds and shut off  accretion \citep[e.g.,][]{Whitworth79,Matzner02,Krumholz06}.  This could halt the evolution of the stellar mass spectrum, perhaps not long after the end of the present simulation. Indeed, as shown in the middle-right panel of Figure \ref{fig:sink}, the most massive sink has an effective temperature of $T_{\rm eff}\approx34,000\,\kelvin$, hot enough to produce ample ionizing luminosity. However, an intense, sustained gas inflow can curtail an H\,II region \citep{Walmsley95,Keto03}. Also, filamentary and disky structures that form in turbulent star-forming environments tend to resist photoionization \citep[e.g.,][]{Peters10, Dale11}. Massive stars preferentially ionize low density gas, allowing the bulk of  accretion to continue through low-filling-factor channels.

\subsection{Protostellar Evolution and the Heating of Dust and Gas}
\label{sec:evolution}

For the first $\approx12$ kyr of sink growth, accretion luminosity dominated protostellar radiative output.  Intrinsic luminosity took over when the first protostar, at $\approx 7\,\msun$, underwent core contraction, 
transitioning from the convective Hayashi track to the radiative Henyey track. Its radius, effective temperature, and intrinsic luminosity increased significantly and abruptly, as is clear in Figure \ref{fig:mesa} and Equation (\ref{eq:l_surf}). This intrinsic brightening is also evident in the middle-right panel of Figure \ref{fig:sink} where the effective temperatures of the most massive sinks rapidly increased from $\approx5000\,\kelvin$ to $\gtrsim25,000\,\kelvin$ in a time span of only $\sim1\,\mathrm{kyr}$. At the end of the simulation, the most massive sink has an effective temperature of $\approx3.3\times10^4\,\kelvin$, hot enough to emit a significant ionizing luminosity, an effect we do not include in this study. 
The total protostellar luminosity is nearly equally divided between the accretion luminosity from all sinks combined and the intrinsic luminosity from the three sinks that have begun hydrogen burning.


\begin{figure}
\includegraphics[width=0.5\textwidth,  trim = 0 30 0 10]{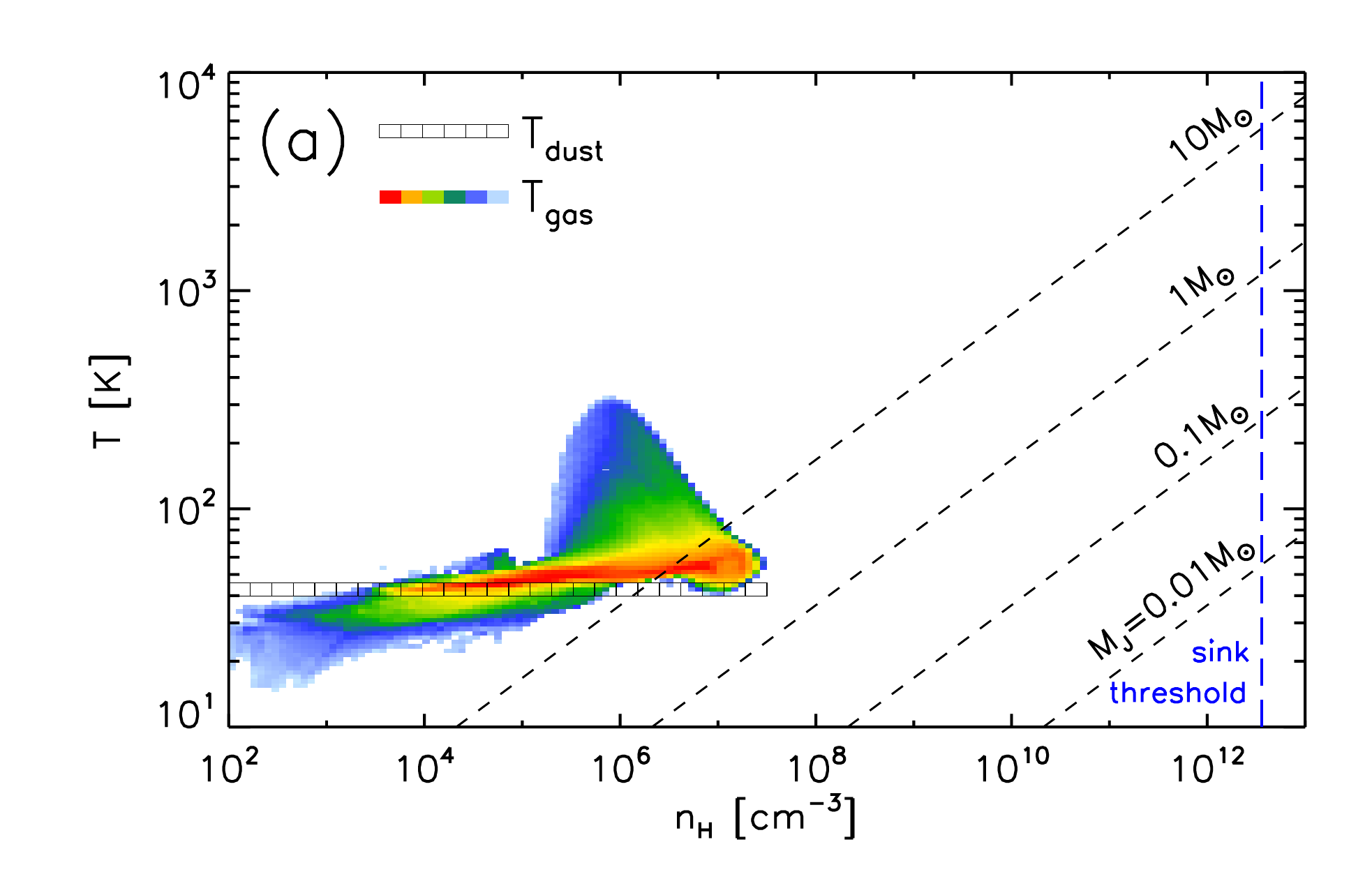} \\ 
\includegraphics[width=0.5\textwidth, trim = 0 30 0 10]{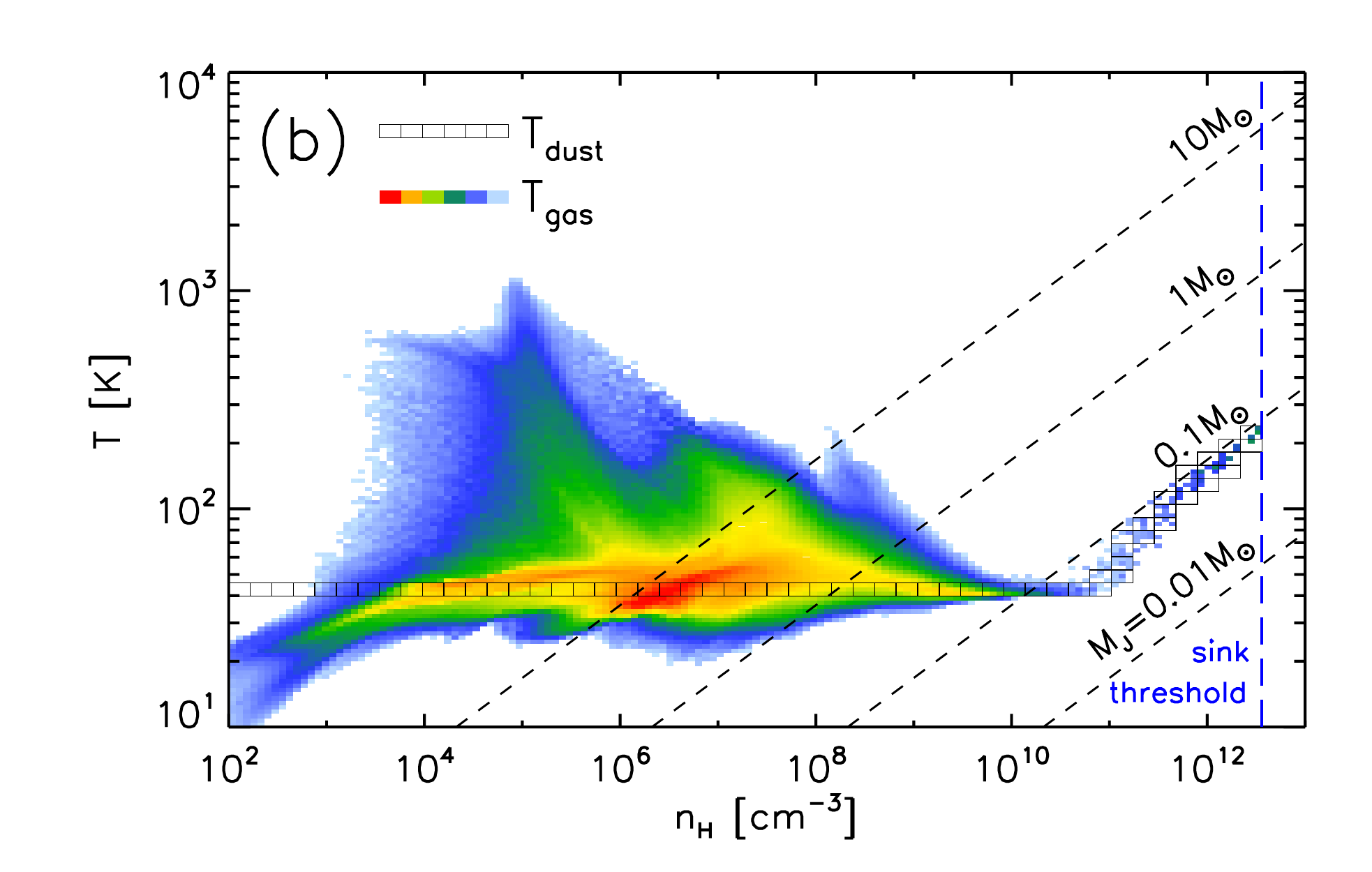} \\ 
\includegraphics[width=0.5\textwidth, trim = 0 30 0 10]{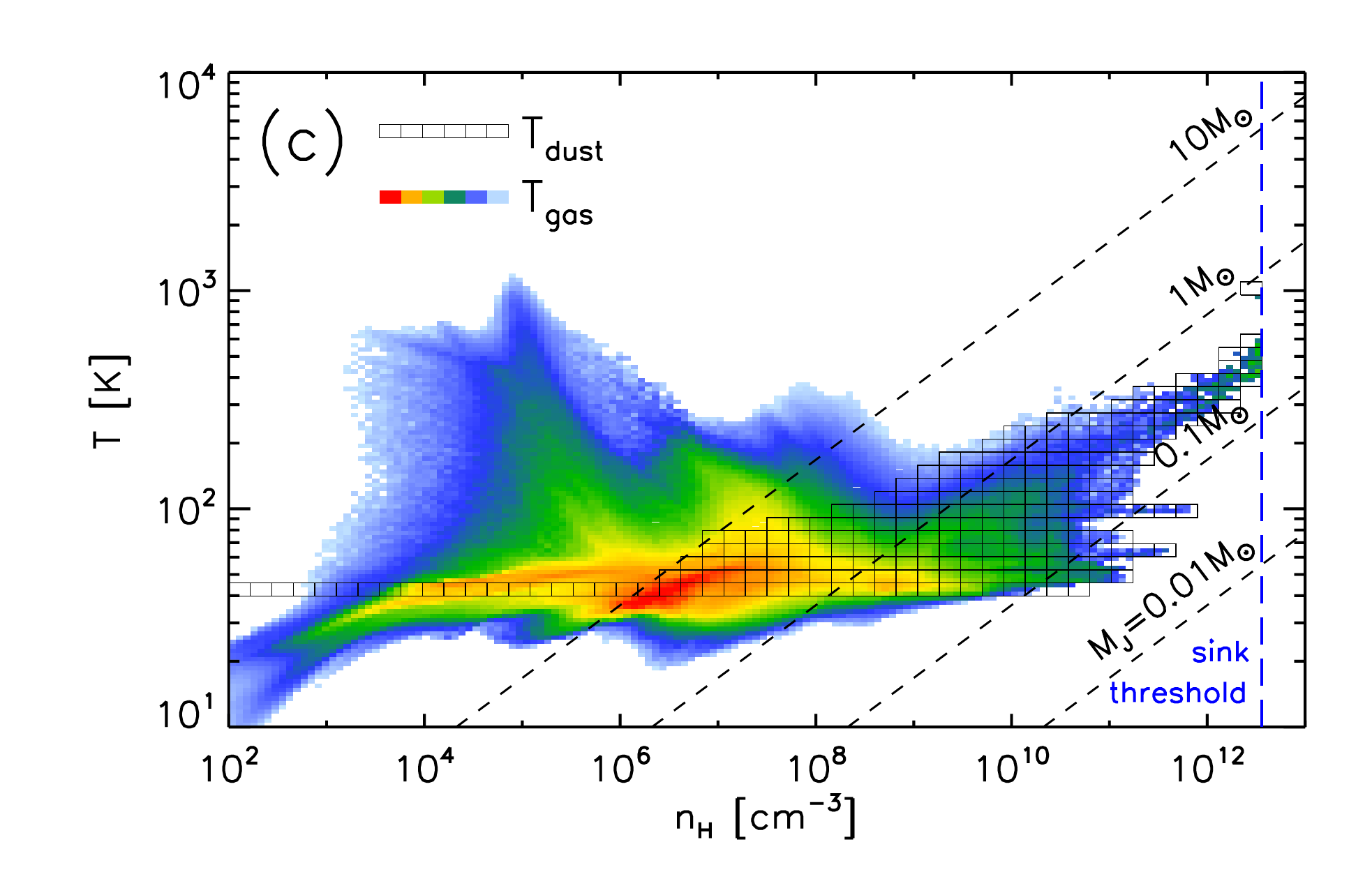} \\ 
\includegraphics[width=0.5\textwidth,  trim = 0 30 0 10]{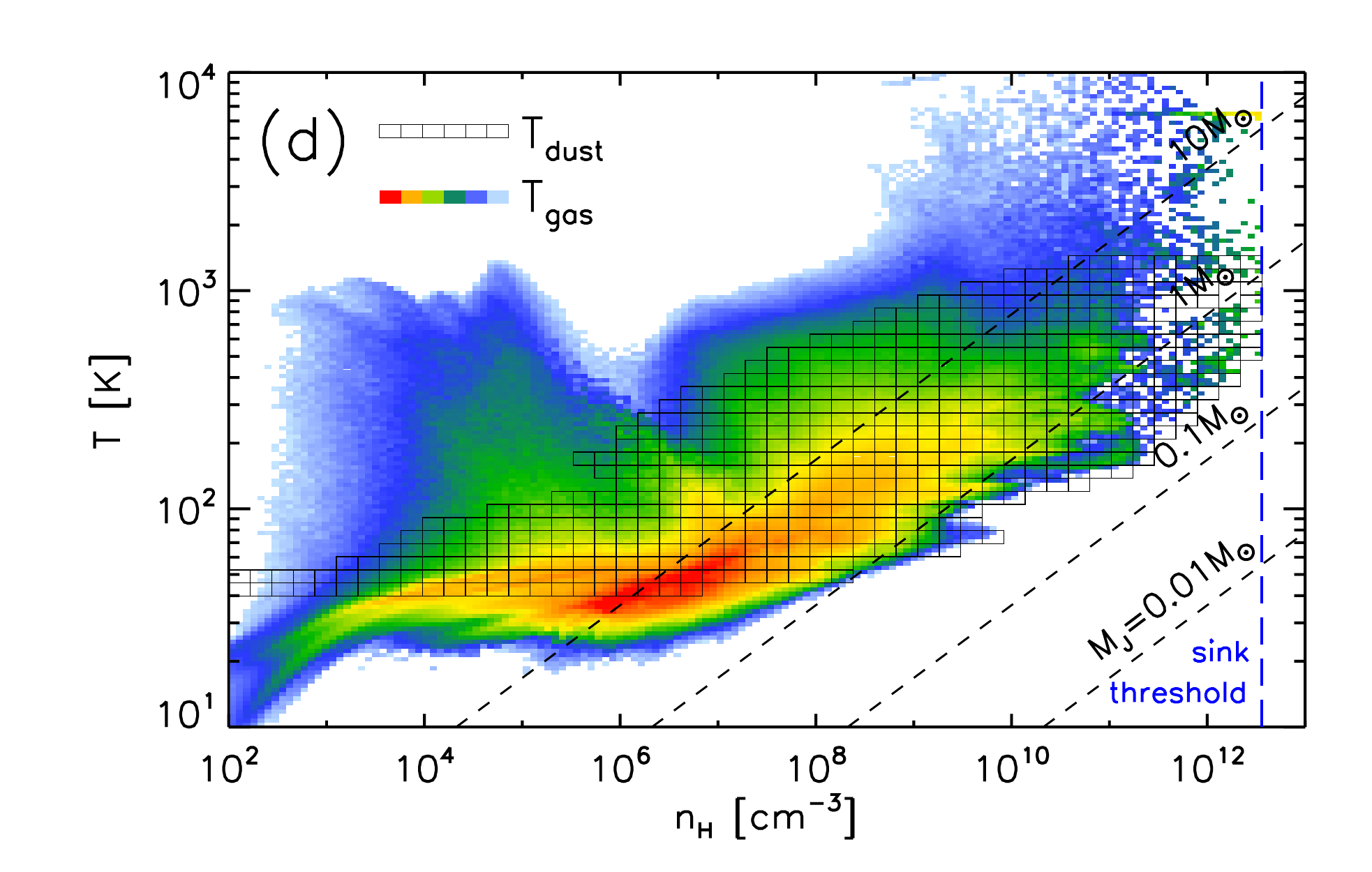}  
\caption{Thermodynamic evolution of the gas and dust. Colored cells represent the amount of gas in density-temperature cells, with red representing the highest gas mass per cell. The dust temperature is overplotted with unfilled rectangles. Dashed lines indicate representative values of the Jeans mass. The panels show the state: (a) $9.6\,\mathrm{kyr}$ after the start of the simulation when the gas begins to depart from isothermal collapse, (b) at the instance of the first sink particle formation, (c) $1.5\,\mathrm{kyr}$ later, when protostellar radiative heating of dust starts to reflect in the gas temperature, and (d) at the end of the simulation. }
\label{fig:dts}
\end{figure}

Thermodynamics modulates fragmentation in the course of gas collapse. In Figure \ref{fig:dts} we explore the thermodynamic evolution in the density-temperature phase plane.  We plot gas and dust temperature against density in four representative simulation snapshots. At the beginning of the excised simulation at redshift $z\approx13.8$, the initial gas temperature equals the CMB temperature  $\tcmb=40\,\kelvin$, while the peak gas density is $\nh\approx10^7\,\cc$. As gas collapse proceeds, two effects act to gradually raise the temperature. First,  [C\,II] and [O\,I] cooling becomes less effective as these lines become optically thick, both intrinsically and to the dust continuum. Second, the gas is slightly heated by the exothermic formation of $\htwo$ on the surfaces of dust grains and to a lesser extent in three-body reactions. This is evident in Figure \ref{fig:dts}a, showing the state of the gas after $9.6\,\mathrm{kyr}$. However, the gas temperature drops back to $\approx \tcmb$ soon thereafter, when at density $\nh\sim10^8\,\cc$, gas becomes thermally coupled to dust. In Figure \ref{fig:dts}b, gas becomes optically thick to the continuum at $\nh\sim10^{11}\,\cc$ and begins to evolve adiabatically until sink formation is possible at $\nh=4\times10^{12}\,\cc$. Protostellar radiative heating of dust grains alters this basic picture somewhat, by increasing the thermal Jeans mass in proximate collapsing cores.  This can be seen in Figure \ref{fig:dts}c recorded after the first sink has grown to $\sim0.5\,\msun$. Finally, Figure \ref{fig:dts}d shows the thermodynamic state at the end of the simulation, $18$ kyr after the first sink formed. The heating of gas via radiative heating of dust is strong at densities $\nh\gtrsim10^7\,\cc$ at which the gas temperature increases with increasing density.

The early thermodynamic evolution of the first protostellar core undergoing gravitational collapse in Figure \ref{fig:dts}a,b fits the predictions of idealized, one-zone models \citep[e.g.,][]{Omukai05,Schneider10,Omukai10}.   By following the gravitational collapse and thermodynamic evolution of gas in one spatial dimension, \citet{Omukai10} estimated that accretion luminosity does not suppress fragmentation for gas metallicities below $10^{-2}\,\zsun$, but only considered the impact on the birth core. Figure \ref{fig:dts}c,d, however, shows that as protostars grow, radiative heating of dust grains produces significant departures from one-zone collapse models \citep[also see][]{Dopcke13}.  Late accretion onto sinks and the formation of new sinks in sites pre-heated by protostellar radiation are significantly modified by dust heating. This highlights the important role that protostellar radiative feedback plays in modulating the star formation process, even at sub-solar metallicities.




\subsection{The Imprinting of a Characteristic Mass}

The thermodynamic behavior discussed in the previous sub-section seems to plays a central role in imprinting a characteristic mass on the fragmentation process. Observations have shown that the characteristic stellar mass scale, i.e., the median stellar mass, is remarkably constant in the Galaxy across a wide range of stellar environments \citep[e.g.,][]{Bastian10}. A median mass that varies with metallicity, redshift, halo mass, or other factors would have implications for any remotely detailed modeling of reionization and chemical enrichment and especially for the interpretation of extragalactic observations. In the present simulation, there is a nearly time-invariant median sink particle mass of $\tilde{M}_*\sim0.5\,\msun$. Sinks with higher masses seem to grow competitively, by accreting from a gas reservoir shared with other sinks.  On the other hand, below the median mass, momentum kicks acquired in close and many-body encounters terminate accretion, leaving the sink with its initial bound core mass (see Fig.\ \ref{fig:sink} and \ref{fig:sink_mf}). What is the physics that selects $\tilde{M}_*$ as the characteristic mass scale? 

Since the equations of ideal, isothermal, self-gravitating (magneto-) hydrodynamics are scale-free \citep[e.g.,][]{Krumholz14}, explanations for a characteristic mass scale require a departure from isothermality.\footnote{Alternatively, a characteristic mass scale could also be imprinted by a fine tuning of initial conditions by a process preceding, and external to star formation.  We do not find this hypothesis attractive.} This departure is generally attributed to either protostellar feedback processes, such as protostellar radiative heating of gas that couples thermodynamics to star formation \citep{Bate09a,Krumholz11}, or chemical, radiative, etc., processes intrinsic to the gas that select specific temperature and density scales \citep[e.g.,][]{Larson03}. In the remainder of this sub-section, we focus on the latter. Idealized, one-zone models for the gravitational collapse and fragmentation of metal-poor clouds have identified the redshift (which sets the CMB temperature floor), gas metallicity, and dust-to-gas ratio as the parameters with the strongest influence on the characteristic fragmentation mass \citep[e.g.,][]{Omukai05,SafranekShrader10,Schneider10}. These studies, however, did not explore many other effects such as radiative feedback, magnetic fields, or supersonic turbulence. 

For the system in our simulation with metallicity $Z=10^{-2}\,\zsun$ at redshift $z=13.8$, one-zone models predict a characteristic fragmentation mass of $\sim1\,\msun$ \citep[e.g.,][]{Omukai05,Schneider10,Schneider12a}.  This mass stems from the onset of dust-gas coupling and departure from isothermality that should set in at densities $\nh\sim10^8\,\cc$, in good agreement with our findings. It supports the hypothesis that dust grains, and not metal fine-structure lines, are responsible for moderating the Pop III to Pop II star formation transition.  Metal line cooling, however, plays a dominant role in regulating fragmentation at more moderate densities $\nh\sim10^{3}-10^{5}\,\cc$ \citep{SafranekShrader14}. 

If indeed the characteristic stellar mass variation can be modeled as in \citet{Schneider10}, the relationship between characteristic mass, redshift, metallicity, and dust content can be used as a probe of ancient stellar populations via dwarf galaxy archaeology \citep[e.g.,][]{Frebel12}. Further hydrodynamic studies surveying a wider parameter space are necessary to validate the applicability of one-zone models to quantifying star formation in realistic, multi-dimensional, turbulent environments.




\section{Discussion and Conclusions}
\label{sec:discussion}

\citet{SafranekShrader14} and the present companion paper taken together trace the process of low-metallicity, primordial star formation from the initial conditions imprinted in cosmic density fluctuations to the formation of individual protostars. In \citet{SafranekShrader14} we simulated a 1 comoving Mpc$^3$ cosmological volume beginning at $z=145$ until a halo with $\tvir=10^4\,\kelvin$ has virialized at $z=13.8$. At that point, the initially metal-free gas was assigned a non-zero metallicity of $10^{-2}\,\zsun$, crudely modeling chemical enrichment by preceding Pop III stars.  The densest, metal-enriched gas cooled isobarically via [C\,II] and [O\,I] line emission to $\tcmb=40\,\kelvin$.  The cold gas fragmented into several few-hundred-solar-mass clumps. The present simulation was initialized by excising \emph{one} of these fragmentary clumps, continuing the simulation to densities $\nh>10^{12}\,\cc$, and inserting sink particles to track individual protostars in the clump. Over the course of $\approx18\,\mathrm{kyr}$ of star formation, $30$ sink particles formed with a combined mass of $81\,\msun$. The individual sink masses ranged from $0.05\,\msun$ to $14.4\,\msun$.  Massive stars had a strong radiative and dynamical impact on the star formation process, stunting the growth of lower mass protostars and suppressing the formation of new ones.

The present simulation overlaps with and extends the early run reported in \citet{SafranekShrader14a}, hence it merits examining any potential differences between the two runs. After $7\,\mathrm{kyr}$ of protostellar formation, the \textsc{heat} simulation of \citet{SafranekShrader14a} formed 37 sinks with a total mass of $15\,\msun$. In the same time period, the present simulation formed only 12 sinks with a total mass of $15\,\msun$. While the star formation efficiencies match, a factor of $\approx3$ fewer sinks in the present simulation deserves scrutiny. Two differences between these otherwise identical simulations contribute to this discrepancy. The present simulation was run at a slightly lower resolution, with a sink particle accretion radius of $\racc=20\,\au$ compared to $10\,\au$ in \citet{SafranekShrader14a}. The new reduced resolution is still sufficient to allow the gas to cross the opacity limit for fragmentation (Equation \ref{eq:n_tau}) preceding sink particle creation. A more probable cause for the discrepancy is different subgrid prescriptions for protostellar evolution, as discussed in Section \ref{sec:protostellar}. Compared with the approximate treatment in \citet{SafranekShrader14a} that assumed quasi-spherical, photosphere-blanketing accretion, the \textsc{mesa} generated protostellar evolutionary models assume thin disk accretion and yield much smaller protostelar radii, and thus much larger accretion luminosities. Indeed, the green curve in the middle-left panel of Figure \ref{fig:sink} shows that if the expression from \citet{Stahler86} (our Eqn.\ \ref{eq:r_stahler}) were used instead of the \textsc{mesa} tracks, the total accretion luminosity would be roughly an order of magnitude lower, except immediately after sink insertion.

The onset of metal line cooling in the parent virialized cosmological object triggered thermal instability that allowed gas to cool quasi-isobarically to the CMB temperature. At this point, fragmentation set in, so in \citet{SafranekShrader14}, \emph{coarse} sink particles were utilized to follow the evolution of the fragments for $\sim4$ Myr. The coarse sinks did not represent individual protostars but instead pre-stellar clumps \citep[e.g.,][]{Bergin07}, each likely to fragment into multiple protostars.  In the cosmological run with a metallicity matching the one in the present simulation, $11$ such clumps formed separated by $100-500\,\mathrm{kyr}$ intervals. We reiterate that the present paper investigates the formation of stars in \emph{one} of these clumps, specifically the first one to form, and follows its internal evolution for $\approx60\,\mathrm{kyr}$. 

If our results are representative of how stars form in all clumps and we can calibrate their star formation efficiencies to that of the first clump, we estimate that our canonical metal-enriched atomic cooling halo forms a stellar mass of $\sim300\,\msun$ in all clumps after $\sim10\,\mathrm{kyr}$ of stellar growth in each clump.  If the entire mass of each pre-stellar clump is converted into stars over a longer period of time, the final stellar mass is $\sim3000\,\msun$. Even the latter, more optimistic estimate of the star formation yield implies a relatively low halo-wide star formation efficiency of $M_{*,{\rm tot}} / M_{\rm b}\sim6\times10^{-4}$, where $M_{*,{\rm tot}}$ is the total stellar mass and $M_{\rm b}$ is the total baryonic mass in the halo.

These calculations allow us to assess the prospect that star clusters of this kind can be detected with next-generation telescopes, such as the JWST. Following \citet{Pawlik11}, we first use the stellar population synthesis models of \citet{Schaerer03} to estimate the initial luminosity $L({\rm H}\alpha)$ in the ${\rm H}\alpha$ line and the luminosity $L({\rm He}1640)$ in the ${\rm He}$\,\textsc{ii}  line, as well as recombination line and the initial ultraviolet (UV) continuum luminosity per unit frequency $L({\rm UV}1500)$ at the restframe wavelength $1500\,\mathrm{\AA}$ as produced by an instantaneous, metallicity $Z=10^{-2}\,\zsun$ starburst forming a $3000\,\msun$ star cluster.  This yields $L({\rm H}\alpha)=3\times10^{38}\,{\rm erg}\, {\rm s}^{-1}$, $L({\rm He}1640) = 6\times10^{35}\,{\rm erg}\, {\rm s}^{-1}$, and $L({\rm UV}1500) = 2\times10^{24}\,{\rm erg}\, {\rm s}^{-1} \,{\rm Hz}^{-1}$. For a spatially unresolved source, we can translate these luminosities into observed flux densities using, e.g., Equations (4) and (5) of \citet{Pawlik11}. For the combined stellar and nebular UV continuum intensities, we find $f_{\nu}({\rm UV}1500) \approx 10^{-3}\,{\rm nJy}$. Similarly, assuming a spectral resolution of $R=\lambda / \Delta\lambda=3000$ for ${\rm H}\alpha$ and $R=1000$ for ${\rm He}1640$, we find $f_{\nu}({\rm H}\alpha) \approx 1\,{\rm nJy}$ and $f_{\nu}({\rm He}1640) \approx 2\times10^{-4}\,{\rm nJy}$. Given a source redshift of $z=13.8$, JWST will detect the redshifted UV continuum at $\lambda_0=2.2\,\microm$ with the NIRCam instrument, ${\rm H}\alpha$ at $\lambda_0=9.8\,\microm$ with the MIRI spectrograph, and the He \textsc{ii} recombination line at $\lambda_0=2.4\,\microm$ with the NIRSpec spectrograph. With a $10^4\,{\rm s}$ exposure, the sensitivities of these three instruments are $10$, $3\times10^3$, and $9\times10^2\,{\rm nJy}$, respectively, to achieve a signal-to-noise ratio of 10, clearly many orders of magnitude higher than the anticipated fluxes.

Even assuming (incorrectly) that the IMF would grow to be top-heavy, which would substantially increase the stellar and nebular luminosities---particularly in the ${\rm He}1640$ line---it is clear that the cluster does not fall within the deep, broadband or spectroscopic detection limits of JWST \citep[also discussed in, e.g.,][]{Gardner06,Johnson09,Pawlik11,Smith14}. The cluster may be magnified by foreground gravitational lenses \citep[e.g.,][]{Zackrisson12,Zackrisson14}, but unfortunately, even extreme magnifications ($\mu\gg100$) are insufficient for a direct JWST detection. The high-redshift, metal-enriched stellar systems that \emph{will} be detected with JWST will not be the very first metal-enriched stellar systems, but more evolved galaxies that have almost certainly been polluted by multiple generations of supernovae. These more evolved systems likely reside in much larger dark matter haloes, with virial masses $M_{\rm vir}\gtrsim10^{9}\,\msun$ \citep[e.g.,][]{Pawlik11}

These considerations indicate that astronomical probes of Pop III and the earliest Pop II star formation must rely on alternatives to direct point source detection at a high redshift. A promising approach involves chemical mapping of metal-poor stars in the Galactic halo and UFD satellite galaxies. Owing to their very low metallicities, chemical heterogeneity, and prevalence of old stars, these systems are considered to be relics of the first galaxies \citep{Ricotti05,Bovill09,Frebel12,Brown13,Brown14}. The stellar system we simulated here shares  physical characteristics with the faintest UFDs---Bo\"otes II, Segue I and II, and Willman I---in terms the  dark matter mass collocated with the stars, the estimated stellar mass, and metallicity.  Simulations tracking the first metals to their progenitor Pop III supernovae, following the subsequent chemical transport at high numerical resolution, and allowing for fine-grained chemical heterogeneity, are necessary to further elucidate the relation of the first metal-enriched star forming systems to UFDs \citep[see][]{Ritter12,Ritter14}.  

Published simulations of the onset of metal-enriched star formation in the first galaxies \citep{Wise08,Greif10,Maio10,Aykutalp11,Wise12,Muratov13,Johnson13a,Jeon14,Jeon15} have yet to resolve the differentiation of gas clumps into individual protostars and could not derive the star formation efficiency or the shape of the IMF from first principles.  Our simulation was specifically designed to return predictions of these key variables.  While a single simulation does not allow us to explore statistical variation, it demonstrates that computationally expensive ``direct", yet fully cosmological, simulations of star formation can be used to calibrate subgrid models that can then be used in coarser ``large-eddy" simulations of star formation in much larger cosmological volumes and across longer redshift intervals.  This is particularly crucial for modeling the early stages of cosmic reionization, currently plagued by substantial uncertainties regarding the star formation efficiency in the smallest galaxies.

\section*{Acknowledgments}

C.~S.~S.\ thanks N.~J.~Evans II, R.~Klessen, M.~Krumholz, A.~Loeb, and many others for useful comments and discussion.
The FLASH code was in part developed by the DOE-supported Flash Center
for Computational Science at the University of Chicago. The authors acknowledge the Texas Advanced Computing Center at The University of Texas at Austin for providing HPC resources under XSEDE allocation TG-AST120024. C.~S.~S.\ is grateful for support provided by the NASA Earth and Space Science Fellowship (NESSF) program. This study was supported in part by NSF grants AST-1009928 and AST-1413501 and NASA grant NNX09AJ33G, to M.~M.\ and V.~B., as well as by  NSF grant AST-1312983 and NASA grant NNX12AC96G to M.~H.~M.

\footnotesize{
\bibliographystyle{mn2e_fixed}
\bibliography{complete}

\IfFileExists{\jobname.bbl}{}
 {\typeout{}
  \typeout{******************************************}
  \typeout{** Please run "bibtex \jobname" to optain}
  \typeout{** the bibliography and then re-run LaTeX}
  \typeout{** twice to fix the references!}
  \typeout{******************************************}
  \typeout{}
 }
}

\label{lastpage}

\end{document}